\documentclass[11pt]{article}

\usepackage[utf8]{inputenc}
\usepackage[T1]{fontenc}
\usepackage{lmodern}
\usepackage{microtype}
\usepackage{amsmath,amssymb}
\usepackage{graphicx}
\usepackage{booktabs}
\usepackage{longtable}
\usepackage{array}
\usepackage{caption}
\usepackage[margin=1in]{geometry}
\usepackage[round]{natbib}
\usepackage[hidelinks]{hyperref}
\usepackage{url}
\graphicspath{{figures/}}

\title{From Stochastic to Stable\\[4pt]
\large Rank Stability and Structural Sufficiency in AI Visibility Measurement}
\author{Ronald Sielinski\\ IQRush\\ \texttt{ron@iqrush.ai}}
\date{}

\begin{document}
\maketitle

\begin{abstract}
AI visibility measurement is fundamentally a comparative exercise: practitioners want to know which domains generative search engines cite most often and whether observed differences are large enough to support competitive decisions. Yet the industry lacks a principled way to determine whether enough data has been collected to support those comparisons. Collection budgets vary widely across studies and platforms, and conclusions are often drawn from rankings whose stability and precision are unknown. We introduce a sequential convergence framework based on two complementary criteria: rank stability evaluates whether the rank-correlation trajectory has reached a structural plateau, and structural sufficiency evaluates whether the spread of citation shares among established domains (those whose citation-share confidence intervals exclude zero) exceeds the uncertainty of those estimates. Together, these criteria distinguish rankings that have merely stabilized from rankings that are sufficiently resolved to support inference. Both criteria are derived from regularities already present in the observed citation distribution, including its rank structure, uncertainty profile, and the boundary between observed and established domains. The framework retains a small number of structural constants but requires no externally specified query count, correlation target, or CI-width target; the stopping decision is driven by directly observed measurement uncertainty and is robust across a range of sufficiency thresholds. Applied across 30 platform-topic combinations spanning Gemini, SearchGPT, and Perplexity, the framework adapts automatically to platform- and topic-specific citation distributions. The results demonstrate that no fixed collection budget can be justified across contexts and that convergence can instead be evaluated from the structure of the observed citation distribution itself. The framework provides a practical basis for determining when AI visibility measurements are ready to support comparative analysis.
\end{abstract}


\medskip
\noindent\textbf{Additional Keywords and Phrases:} generative search, generative search optimization, answer engine optimization, AI visibility.

\medskip

\hypertarget{introduction}{%
\section{Introduction}\label{introduction}}

The purpose of AI visibility measurement is comparative analysis.
Practitioners want to know which domains a generative search engine
cites most often, how that standing changes over time, and whether
observed differences are large enough to matter for competitive
strategy. The goal is not the measurement itself, but the decisions that
it is meant to enable.

The industry quickly converged on a multi-query process for sampling
multiple aspects of a topic and measuring how often a brand's web pages
are cited (or its name is mentioned) in generative search responses. But
the industry has paid far less attention to a critical question: whether
the resulting measurements yield enough data to support the comparisons
they are used to make. Collection budgets differ by nearly an order of
magnitude across published studies and industry practice, with no
principled basis for choosing one over another. Comparative conclusions
are routinely drawn from measurements that may not be adequate to
support them.

The core difficulty is that citation share is a sample statistic, not a
fixed quantity. Generative search engines are stochastic: the same query
submitted on different occasions can produce different responses and
cite different sources. A multi-query design is intended to average over
prompt-level idiosyncrasies, but it also aggregates many variable
outputs, so repeating the same measurement process will likely produce
different results even when nothing meaningful has changed. Citation
counts and shares are therefore random variables, and the rank ordering
derived from them inherits the same variability. When adjacent domains
have similar citation metrics, their relative ranks may remain
statistically unresolved even after substantial data collection. Our
prior work \citep{sielinski2026} established these properties empirically.
The question this paper addresses is what follows from them: how much
data must be collected before rank-based conclusions can be trusted?

The answer requires two conditions to hold simultaneously. The first is
rank stability: the rank ordering must have settled. During early
collection, the ordering shifts substantially with each additional
response, because the domains that will eventually become top-cited have
not yet established consistent leads. Because generative engines cite
only a handful of sources per response---SearchGPT, for example, cites
between 4 and 6---the eventual leaders may not appear at all in the
first several responses. Stability emerges gradually as cumulative
citation counts begin to reflect the underlying distribution.

The second condition is sufficiency: even after rank order has
stabilized, the measurement must be precise enough to support the
comparisons it is used to make. A stable rank ordering can still be too
imprecise to act on---if the domain ranked first and the domain ranked
second have nearly identical underlying shares, the ordering does not
reliably distinguish winners from runners-up. Sufficiency requires that
the spread of citation shares among the reliably present domains exceed
the typical measurement uncertainty---that the signal be at least as
large as the noise.

Both conditions depend on the structure of the underlying citation
distribution, which varies substantially across platforms and topics.
Citation density differs by more than sixfold between platforms; the
concentration of the distribution and the rate at which rank order
stabilizes follow suit. No fixed collection budget can be justified: the
amount of data required depends on the structure of the citation
distribution, not on a schedule set in advance. A fixed rank-correlation
threshold fails for the same reason: convergence response orders under a
fixed \emph{$\rho$}~$\ge$~0.95 criterion range from 40 responses to never across
the 30 platform-topic combinations studied here, and no platform
converges at a rate that is consistently fastest or slowest.

The framework this paper introduces avoids external calibration by
deriving both criteria directly from the structure already present in
the data. Neither criterion requires the analyst to specify a target
rank correlation, a desired confidence interval width, or a
predetermined query count. Rank stability is detected as a structural
plateau in the rank correlation trajectory, identified by model
selection rather than level crossing. Sufficiency is established when
the spread of established-domain shares exceeds the mean measurement
uncertainty---a ratio whose threshold follows from the geometry of
signal and noise rather than from an external specification. The same
criterion automatically demands more data wherever citations accrue
slowly or established shares are tightly packed.

The key to this self-calibration is a consistent structural property of
generative citation distributions. Citation distributions in generative
search engines are broadly consistent with a power-law form: a small
number of highly cited domains accumulate citations at an elevated rate
while the great majority appear sporadically. This pattern holds across
all platforms and topics studied here; what varies are the
parameters---citation density, power-law exponent, noise boundary,
convergence rate. Because both criteria are expressed in terms of
observed distributional features---rank trajectory, citation density,
share spread, and empirical uncertainty---the framework calibrates
itself to the citation distribution it encounters.

The questions this paper addresses fall within the scope of sequential
analysis, the branch of statistics concerned with data collection
protocols that adapt based on accumulated evidence \citep{wald1947}.
Sequential estimation treats sample size as a random outcome rather than
a fixed input, stopping when a specified precision criterion has been
met. The convergence framework developed here instantiates this logic
for citation visibility measurement.

The remainder of this paper is organized as follows. Section~2 provides
background on citation visibility metrics as random variables, the
power-law structure of citation distributions, and the two
manifestations of rank uncertainty that motivate the convergence
framework, and closes with a review of related work. Section~3 describes
the dataset and methods, including bootstrap CI estimation, power-law
segmentation, the rank stability criterion, and the structural
signal-to-noise ratio that operationalizes sufficiency. Section~4
presents results across 30 platform-topic combinations: distributional
structure, rank stability trajectories and convergence response orders,
rank confidence interval widths at full sample size, the conjunctive
convergence results, and a validation analysis comparing stopped
orderings to end-of-window orderings. Section~5 discusses implications
and limitations, and Section~6 concludes.

\hypertarget{background}{%
\section{Background}\label{background}}

\hypertarget{citation-visibility-metrics-as-random-variables}{%
\subsection{Citation Visibility Metrics as Random
Variables}\label{citation-visibility-metrics-as-random-variables}}

In our prior work, we established empirically that citation visibility
metrics are random variables, not fixed values \citep{sielinski2026}. The
same queries submitted to the same platform on different occasions
produce different cited source sets: platform-level Jaccard similarity
across repeated runs of the same query is approximately 0.30 for Gemini,
0.40 for SearchGPT, and 0.50 for Perplexity, meaning that even the most
consistent platform shares only half its cited domains across repeated
samples. This is a structural property of generative systems that use
stochastic sampling at generation time, not measurement error to be
corrected.

The practical consequence is that the citation share for a domain,
defined as the domain's citation count divided by the total citations
across all responses, is a sample statistic with quantifiable
uncertainty. Its variance under a multinomial sampling model is
approximately \emph{$\theta$}(1$-$\emph{$\theta$})/\emph{M}, where \emph{$\theta$} is the true
underlying citation probability and \emph{M} is the total citation
count, so CI width scales as 1/$\sqrt{\phantom{x}}$\emph{M}. Each doubling of data yields
only approximately 29\% improvement in CI width. The law of diminishing
returns applies directly.

Bootstrap confidence intervals, computed by resampling at the response
level and recomputing share for each replicate, make this uncertainty
explicit. Our prior work found that these intervals are wide relative to
the differences between domains: for most frequently cited domains on
SearchGPT, 95\% CI spans range from 3 to 6 percentage points, and
apparent differences between many adjacent domains are statistically
indistinguishable from sampling noise. A domain reported at 9.5\%
citation share and another at 6.0\% may have confidence intervals that
overlap substantially, so the data cannot rule out that the second
domain's true share exceeds the first's.

This finding has a direct implication for rank that the prior paper did
not explore. If the share estimates underlying a rank ordering carry
uncertainty wide enough to make adjacent domains indistinguishable, then
the rank ordering itself carries uncertainty that a single point
estimate of rank cannot capture. Domain A may be ranked first and domain
B second in a single sample, while the bootstrap distributions of their
shares overlap substantially. The rank is real, but the uncertainty
around it is also real. In a repeated sample, B could easily rank above
A. Ignoring that uncertainty produces the same false precision as
reporting share without a confidence interval.

\hypertarget{two-manifestations-of-rank-uncertainty}{%
\subsection{Two Manifestations of Rank
Uncertainty}\label{two-manifestations-of-rank-uncertainty}}

The tension between rank's utility and its uncertainty manifests in two
distinct ways, corresponding to two phases of measurement.

The first is rank instability during collection. The domains that will
ultimately be most frequently cited need time to accumulate enough
citations to establish themselves above the competition. During early
collection, the first 20 to 60 queries, depending on topic and platform,
the rank ordering changes substantially with each additional response,
and the confidence intervals of the top-cited domains overlap broadly
enough that many apparent rank differences are statistically
indistinguishable. This is a structural feature of how citations
accumulate, not a transient artifact: top-cited domains establish
themselves faster than mid-tier domains, but not instantaneously.
Reporting ranks from an insufficient sample produces a misleading
picture of competitive standing, because the ordering has not yet
stabilized.

The second is rank uncertainty in a stable distribution. Even after rank
order has stabilized, adjacent domains in the dense middle of the
distribution can exchange positions across successive samples without
either showing a statistically significant change in share. The
distinction matters: stability concerns whether the ordering is still
changing over collection; uncertainty concerns whether it is precise
enough at a fixed collection size to support inference. A domain that
moves from rank 10 to rank 20 between two measurement periods may have
undergone a meaningful competitive shift, or it may be oscillating
within the band of sampling noise that is indistinguishable from
stability at its share level. Distinguishing these cases requires tools
that go beyond a point estimate of rank and beyond a share-based
significance test.

\begin{figure}[htbp]
\centering
\includegraphics[width=\textwidth,height=0.9\textheight,keepaspectratio]{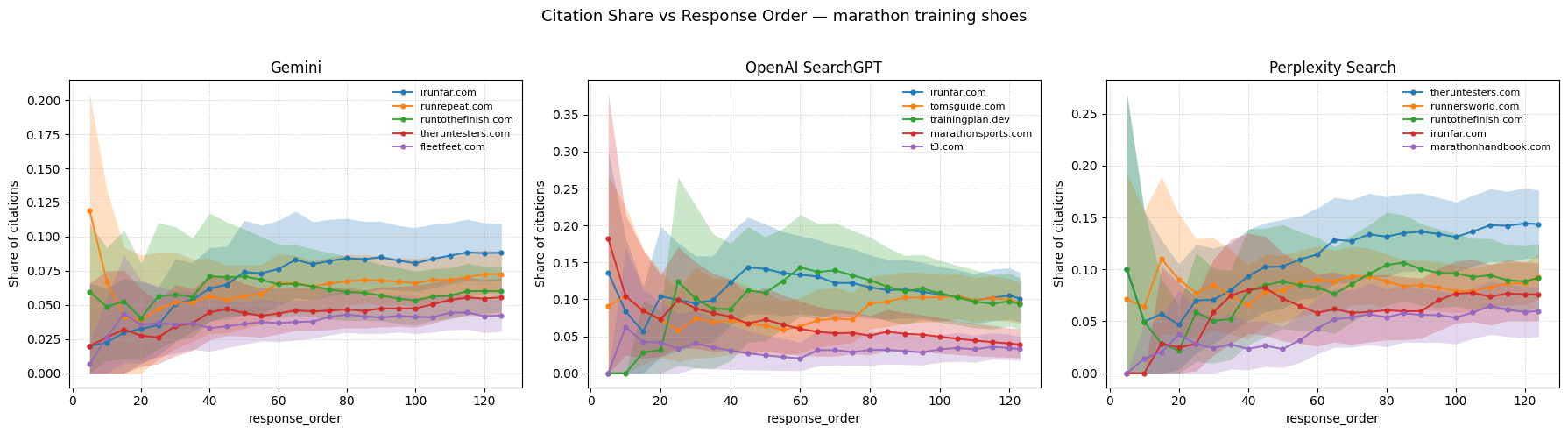}
\caption{Citation share accumulation over response order for the five
domains that finish with the highest citation share at end of
collection. Each line tracks a domain's cumulative citation share as
responses accumulate; the shaded bands represent the domain's share CI.
Histories are shown only for the domains that eventually rank in the top
five; earlier in collection these domains may have been ranked below
their final position, which is not shown.}
\label{fig:1}
\end{figure}

Figure~\ref{fig:1} illustrates three patterns that appear consistently across
topics and platforms. Rank positions among the top-five domains shift
frequently during the first 20 to 40 responses on every platform---the
early citation landscape is volatile. Even after rank order stabilizes,
confidence intervals for individually ranked domains continue to
overlap, so platform preference among similarly-ranked domains remains
statistically ambiguous. Measurable uncertainty persists even in the
most stable configurations: observed ranks are point estimates, not
fixed readings.

\hypertarget{power-law-structure-and-what-it-means-for-rank}{%
\subsection{Power-Law Structure and What It Means for
Rank}\label{power-law-structure-and-what-it-means-for-rank}}

Citation count distributions follow an approximate power-law form, the
rank-frequency counterpart of Zipf's Law \citep{zipf1949}, although the
fitted parameters vary by platform and topic. On log-log axes, citation
counts decline roughly linearly with rank: a small number of highly
cited domains account for a disproportionate share of citations,
followed by progressively less-cited domains down the ordering. Our
prior work documented this structure across multiple platforms and found
the overall shape to be stable across repeated samples, even when
individual domain positions changed between draws. The existence of
sampling variability therefore does not imply the absence of structure;
rather, the rank fluctuations of the preceding section occur within an
underlying distributional form that is itself comparatively consistent.

What matters here is not the internal structure of any single fit but
the variation across fits. The fitted exponent and the citation density
differ substantially across platform-topic combinations, and those
differences are what make the recurring pattern of the preceding section
(early rank instability, then persistent confidence-interval overlap
even among top-ranked domains) resolve at different rates. The power-law
fit therefore serves as descriptive background that quantifies how
citations concentrate across platforms and topics; it is not the basis
for any stopping decision. The convergence criterion derives the
established domain set directly from bootstrap confidence intervals
rather than from a fitted boundary (\S3.4), and the variation in fits is
precisely why no single collection budget transfers across contexts.

\hypertarget{contributions-of-this-paper}{%
\subsection{Contributions of This
Paper}\label{contributions-of-this-paper}}

This paper develops a principled framework for determining when enough
data has been collected for rank-based citation visibility to be
trustworthy. We make three contributions. First, we establish rank
stability during collection as a convergence criterion: using weighted
Spearman rank correlation tracked across response order, we show when
the citation rank ordering has stabilized sufficiently to support
reporting and characterize how that threshold varies by platform and
topic. Second, we show that even after rank order stabilizes, confidence
intervals around rank positions remain wide relative to the differences
between adjacent domains; because CI width shrinks only with the square
root of sample size, continued collection closes that gap slowly and at
diminishing returns. Third, we introduce structural sufficiency as a
principled stopping criterion: a signal-to-noise ratio that compares the
spread of citation shares across the established domain set (domains
whose citation-share CI excludes zero) to their mean CI half-width,
providing a data-driven answer to the cost-benefit tradeoff between
collection effort and inferential precision.

\hypertarget{related-work}{%
\subsection{Related Work}\label{related-work}}

\emph{Stochastic citation systems.} Citation visibility in generative
search has attracted growing empirical attention as answer engines have
displaced traditional search results for many queries. \citet{gao2023}
established that LLM-generated answers can be evaluated for citation
quality independently of generation quality, formalizing the
grounded-generation task that underlies any systematic citation
measurement. Empirical audits have documented that citation patterns are
stochastic, platform-specific, and concentrated in ways traditional
search metrics do not predict: \citet{li2024} found geographic
and commercial bias in citation patterns across ChatGPT, Bing Chat, and
Perplexity across repeated sampling, and \citet{zhang2025} showed that
citation distributions in LLM-based search engines are substantially
more concentrated than those of traditional search, with 37\% of cited
domains unique to LLM-based systems. That concentration pattern directly
motivates the distributional analysis in \S3.4: rank comparisons carry
meaning only over the subset of domains that appear reliably, rather
than those observed only incidentally. \citet{mokander2023} argue that
application-level LLM audits must be systematic and reproducible, a
requirement the convergence criterion developed here is designed to
satisfy.

\emph{Rank stability and uncertainty.} The use of rank as a visibility
metric connects this work to a long tradition in information retrieval.
\citet{jarvelin2002} established that rank position carries
graded importance, the foundational motivation for any rank-based
visibility metric, and \citet{moffat2008} showed that rank-based
retrieval metrics must account for experimental uncertainty and the
elevated importance of top positions. \citet{webber2010} introduced
rank-biased overlap (RBO), a probabilistic similarity measure for ranked
lists that weights top-rank agreement and accommodates non-conjoint
lists, which is relevant here because different platforms frequently
cite different domains entirely. The rank stability criterion developed
in this paper is methodologically adjacent: it uses weighted Spearman
correlation rather than RBO, but similarly concentrates the stability
signal at the top of the distribution. \citet{vigna2015} generalized weighted
rank correlation to handle ties, a necessary extension when many domains
share identical or near-identical citation counts, as is common in the
citation tail.

\emph{Sequential stopping rules.} The problem of determining when enough
data has been collected falls within sequential analysis, where sample
size is treated as a random outcome determined by accumulated evidence
rather than a fixed input \citep{wald1947}. \citet{chow1965}
formalized the fixed-width sequential confidence interval, which stops
when a CI has narrowed to a specified target width, the direct
methodological antecedent of the sufficiency criterion developed here.
Neither the classical sequential analysis literature nor the retrieval
evaluation literature has addressed the convergence problem specific to
citation-share distributions: the joint determination of rank stability
and measurement precision in a stochastic, power-law distributed system
with substantial between-platform variation in effective sample size.
More broadly, both criteria are calibrated to the data's own structure
rather than to externally specified targets: the plateau detector
requires no target correlation level, and the SNR criterion requires no
target CI width. The framework discovers whether enough data has been
collected from the properties of the data itself.

\hypertarget{methods}{%
\section{Methods}\label{methods}}

\hypertarget{data}{%
\subsection{Data}\label{data}}

The analyses reported here use a new dataset collected under the same
protocol as our prior paper \citep{sielinski2026}. The dataset is based on a
set of ten topics that span a range of market segments: consumer goods,
travel, and medical devices, among others. For each topic, 125 queries
were generated by prompting an LLM (ChatGPT) to produce queries
conditioned on a randomly selected query type drawn from a fixed set of
twenty categories: strengths and weaknesses, popular options, emerging
trends, expert recommendations, feature comparisons, and others. The
resulting query sets were submitted to Perplexity Search, OpenAI
SearchGPT, and Google Gemini, and citation extraction followed
platform-specific procedures. The resulting records link each cited URL
to its registered domain, query identifier, response identifier,
platform, and topic. All analyses operate at the domain level. The
dataset structure is summarized in Table~\ref{tab:1}.

\hypertarget{bootstrap-resampling-and-citation-share-estimation}{%
\subsection{Bootstrap Resampling and Citation Share
Estimation}\label{bootstrap-resampling-and-citation-share-estimation}}

Citation visibility metrics are sample statistics computed from a finite
collection of responses and are therefore subject to sampling
variability. We characterize this variability using multinomial
bootstrap resampling at the query level rather than the citation level,
which correctly propagates within-response citation clustering:
citations within a single response are a joint draw, not independent
observations, and treating them as exchangeable understates uncertainty
by assuming independence that does not hold.

The bootstrap procedure resamples queries with replacement at each
iteration, recomputing citation share as the domain's citation count
divided by total citations across the resampled responses. Citation
share for domain \emph{d} within platform \emph{p} at bootstrap
iteration \emph{b} is:
\[{\hat{s}}_{d,p}^{(b)}\text{=}\frac{c_{d,p}^{(b)}}{\sum_{d^{'}}^{}c_{d^{'},p}^{(b)}}\]
where both numerator and denominator are recomputed from each bootstrap
draw, correctly propagating the dependence of share on total citation
volume. Confidence intervals used in reporting are computed using the
BCa (bias-corrected and accelerated) method throughout \citep{efron1987},
which adjusts for the systematic downward bias in bootstrap
distributions for sparse citation data. All bootstrap confidence
intervals reported in this paper use B = 1,000 replicates. Doubling B
from 500 to 1,000 moves roughly half the sufficiency firing orders later
by one to two responses, and the rank-stability test involves no
resampling at all, so the stopping decisions are insensitive to the
replicate budget at this scale. BCa applies to the share-level intervals
that define the established tier and gate the sufficiency criterion;
derived statistics, including the rank confidence intervals of \S4.3 and
the share-trajectory bands, use bias-corrected percentile intervals,
because the acceleration term does not transfer through post-transforms.
All resampling draws whole queries within a platform-topic bucket, which
conditions every interval on citation-bearing responses; the rank
confidence intervals additionally share one query draw across platforms,
a paired design stratified by topic.

\hypertarget{reliability-of-bootstrap-confidence-intervals}{%
\subsection{Reliability of Bootstrap Confidence
Intervals}\label{reliability-of-bootstrap-confidence-intervals}}

Bootstrap confidence intervals on citation share are not uniformly
reliable across the citation count distribution, and rank confidence
intervals inherit the same structure. The cause is within-query citation
clustering: a domain cited five times within a single query behaves very
differently under bootstrap resampling than one cited once across five
queries, despite identical total counts. The former contributes its
entire citation mass only when that query is resampled, producing a
zero-inflated distribution that percentile intervals understate, and the
BCa correction addresses this only partially in the sparse regime. The
practical consequence is a danger zone at roughly \emph{k} = 5--20
citations, where interval widths appear stable while the underlying
distributions remain skewed (Gemini reaches distributional symmetry near
\emph{k} $\approx$ 20 citations; SearchGPT does not within the observed range).
Because rank is computed from share, rank confidence intervals carry the
same caveat, which we flag when interpreting them in \S4.3.

\hypertarget{observed-and-established-domains}{%
\subsection{Observed and Established
Domains}\label{observed-and-established-domains}}

\begin{figure}[htbp]
\centering
\includegraphics[width=\textwidth,height=0.9\textheight,keepaspectratio]{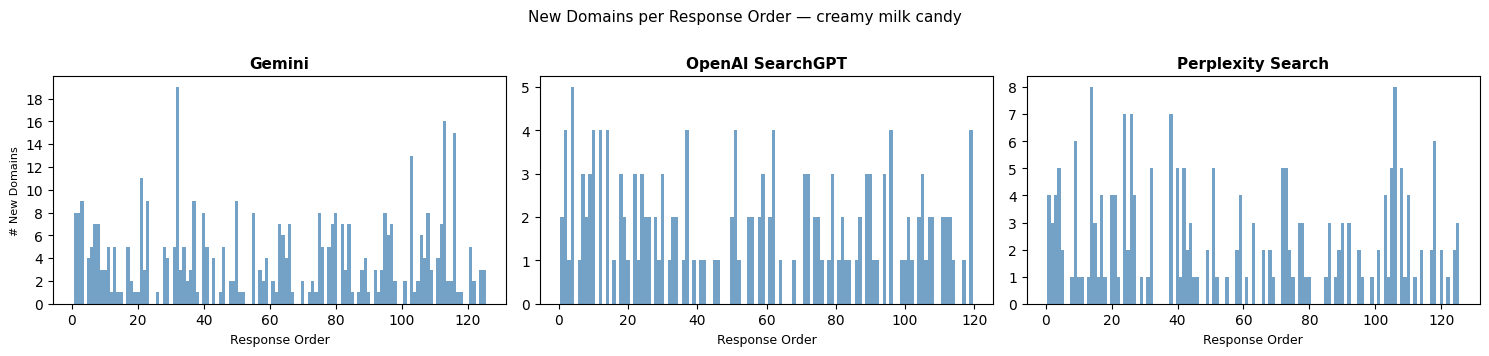}
\caption{The number of new (previously unseen) domains per response
order for the ``creamy milk candy'' topic, all three platforms.}
\label{fig:2}
\end{figure}

It is natural to expect that the domains a platform cites most often
will be the first to surface, so that the leaders announce themselves
early in collection. They do tend to appear sooner than tail domains,
but only on average: because each response cites just a handful of
sources, the eventual leaders can be missing from the first several
responses, and the early ordering is correspondingly noisy. A second and
more consequential expectation is that new domains will eventually stop
appearing, which would hand the analyst a natural signal to stop
collecting once the vocabulary has settled. They do not. Figure~\ref{fig:2} shows
previously unseen domains entering the result set throughout the window,
at a decelerating but non-vanishing rate.

This continued growth is what the distributional structure predicts. The
vocabulary follows Heaps' Law (Heaps, 1978): the number of distinct
domains grows as a concave power of collection size, so new domains keep
arriving and never stop within any feasible window. Its companion,
Zipf's Law \citep{zipf1949}, the rank-frequency form of the same power law
(\S2.3), explains what those newcomers are: a few domains accumulate
citations rapidly while the great majority are cited only a handful of
times, many of them just once. Heaps' Law guarantees the stream of new
domains; Zipf's Law guarantees that most of them never amount to
anything. Waiting for the vocabulary to stop growing is therefore not a
usable stopping rule.

The problem this creates is one of distinction: most domains that appear
are noise, seen once or twice and unlikely to recur, while a minority
are reliably present, and a convergence criterion must operate on the
latter. Within the observed citation vocabulary we therefore separate
two nested tiers by statistical detectability. A domain is observed once
it has been cited at least once. A domain is established when its
citation-share confidence interval excludes zero, that is, when its
observed presence is robust to response-level resampling of the query
set. Observed domains include everything seen in the collection;
established domains are the subset whose presence is reliable enough to
persist across repeated runs of the same analysis. The established tier
is the domain set over which structural SNR is computed and the primary
reporting-relevant rank uncertainty is evaluated.

The boundary is set directly by the data: a domain becomes established
when its accumulated citations are enough for a bootstrap confidence
interval to clear zero. Because BCa intervals correct for the
zero-inflation and asymmetry of sparse bootstrap distributions, the
boundary adapts automatically to within-query clustering and citation
density. No minimum domain count is required, and no power-law fit is
needed to define membership; the framework does not depend on fitting
the distribution at all, and the fits reported in \S4.1 are descriptive.
Which observed domains will ever become established cannot be known in
advance, and the established tier grows more slowly than the full
vocabulary as that status emerges from the data.

\hypertarget{rank-stability-during-collection}{%
\subsection{Rank Stability During
Collection}\label{rank-stability-during-collection}}

Rank order is not immediately interpretable from the first query. During
early collection, the domains that will eventually become the most
frequently cited have not yet accumulated enough citations to establish
themselves: their confidence intervals overlap broadly with mid-tier and
even lower-tier domains, and the rank ordering changes substantially as
each additional query is processed. This is not a transient artifact but
a structural property of Zipfian processes: the top-cited domains need
time to separate from the field.

We assess whether rank order has stabilized using the weighted Spearman
rank correlation between successive collection windows. Successive
windows overlap---each contains the observations of the previous window
plus the new increment---so the correlation trajectory is strongly
autocorrelated and, by construction, slow-moving:
\[\rho_{t}\text{\ =\ }\rho_{\text{Spearman}}\left( {\hat{\theta}}_{t\text{-}w},\text{\ }{\hat{\theta}}_{t} \right),\text{\ \ }w_{d}\text{=\ }\frac{{\hat{\theta}}_{d,t\text{-}w}\text{+}{\hat{\theta}}_{d,t}}{2}\]
Weights are the average shares across the two windows, so rank swaps
among high-share domains contribute more to the correlation than swaps
in the low-share tail, where single-citation fluctuations routinely
reorder positions without competitive meaning. The statistic is
deliberately computed over the full observed vocabulary rather than the
established tier alone: the share weights already place roughly three
quarters of its mass on established domains at end of window (0.42 to
0.92 across the 30 combinations), and leaving the universe unrestricted
keeps rank stability free of the bootstrap machinery that defines
establishment, so the two convergence conditions supply independent
evidence. The residual gap between full-ordering stability and
established-set precision is precisely what the sufficiency condition
and the validation analysis of \S4.6 measure. The two collections
compared at response order \emph{n} are the cumulative samples through
response \emph{n} and through response \emph{n} $-$ \emph{w}, where
\emph{w} = max(3, $\lceil$0.10 $\cdot$ \emph{n}$\rceil$).

The proportional lag counteracts a mechanical drift in \emph{$\rho$} that
would otherwise appear as \emph{n} grows. At \emph{n} = 10, a single new
response shifts cumulative citation shares by roughly 10\%; at \emph{n}
= 100 the same response moves them by 1\%. Consecutive snapshots
therefore grow more similar as \emph{n} increases, pushing \emph{$\rho$}
toward 1 even when ranks are still settling. Stretching the lag in
proportion to \emph{n} holds the share-movement gap between the two
snapshots roughly constant, so the correlation trajectory remains
informative at every sample size.

This construction gives the statistic a natural form of smoothing that
requires no explicit filter. Because the windows are cumulative, the
increment from one response order to the next is a fraction on the order
of 1/\emph{n} of the accumulated counts, so each additional response
perturbs the ordering less than the one before, and the trajectory is
mechanically damped as collection proceeds---once an initial warm-up
period has passed and the established set has accumulated sufficient
citations for the 1/\emph{n} dampening to dominate window-to-window
noise. The share weighting compounds this: the top-cited domains both
anchor the correlation and extend their lead fastest, so once a top
domain has accumulated a substantial count it becomes increasingly
difficult to displace, and the weighted correlation it dominates settles
even while the long tail keeps churning. The trajectory therefore rises
and plateaus rather than oscillating freely, which is what makes a
plateau---rather than a particular level---the natural object to detect,
and what lets the detector below operate on the raw correlation without
a separate smoothing step. The weighted correlation tends to increase as
collection proceeds, although local declines remain possible.

A natural first approach might be to declare rank order stable when the
correlation crosses a fixed threshold (e.g., \emph{$\rho$} $\ge$ 0.95). The
results of this paper, however, expose a structural problem with any
fixed level: citation density determines the ceiling \emph{$\rho$} can reach,
regardless of whether the rank order has stopped changing. A sparse
platform's correlation may plateau well below 0.95 even after the
citation distribution has stabilized, because sampling noise between
successive collection windows suppresses the observed rank correlation
even when the true distribution is no longer changing.

\begin{figure}[htbp]
\centering
\includegraphics[width=\textwidth,height=0.9\textheight,keepaspectratio]{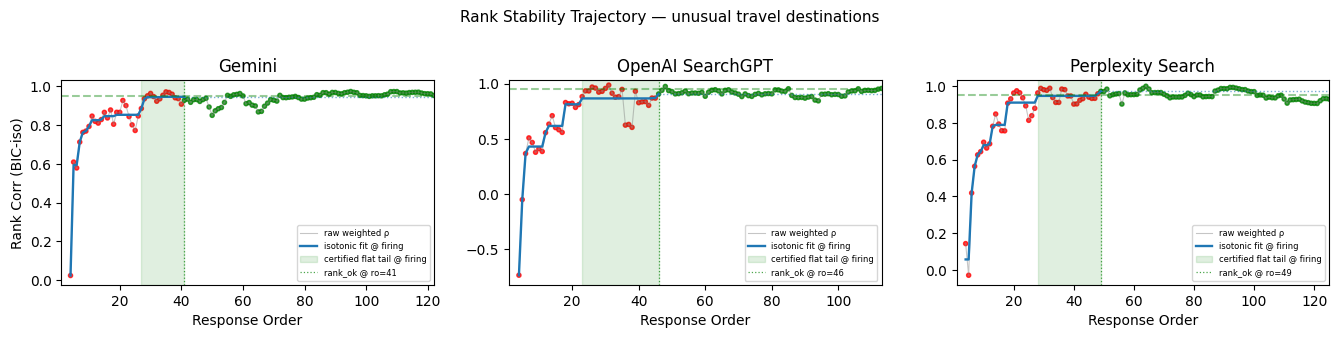}
\caption{Weighted Spearman rank correlation versus response order for
the ``travel destinations'' topic, shown separately for Gemini,
SearchGPT, and Perplexity. Each point represents the raw weighted
\emph{$\rho$} . The points are red prior to rank\_corr\_ok and green
thereafter. The BIC-iso plateau fit is overlaid on each trajectory; the
solid blue line illustrates the isotonic fit, and the green vertical
band identifies the BIC-chosen flat tail. The dashed horizontal line
marks the fixed stability threshold (\emph{$\rho$} = 0.95). Table~\ref{tab:3} reports
convergence orders for all 30 platform-topic combinations.}
\label{fig:7}
\end{figure}

The criterion this paper uses detects that plateau directly with a
segmented model fit to the raw correlation trajectory (Figure~\ref{fig:7}, in the
Results section, shows the full collection of trajectories to which this
test is applied). The rising phase is fit by isotonic (monotone
non-decreasing) regression via the pool-adjacent-violators algorithm
(PAVA; Barlow et al., 1972). Because the trajectory rises toward a
ceiling, the isotonic fit absorbs the window-to-window noise without a
bandwidth or smoothing constant and, unlike a straight line, does not
misrepresent the concave, Heaps- and Zipf-shaped approach to that
ceiling. A flat (constant) tail is then appended at a changepoint; both
the changepoint location and whether a changepoint is warranted at all
are selected by the Bayesian information criterion against a pure-rise
model, and the tail must in addition be better described as flat than as
linear under an effective sample size \emph{n}\textsubscript{eff} =
\emph{n}(1 $-$ \emph{$\rho$}\textsubscript{1})/(1 + \emph{$\rho$}\textsubscript{1}), where \emph{$\rho$}\textsubscript{1} is the lag-1
autocorrelation of the linear-fit residuals \citep{bartlett1946}. Rank
stability (rank\_corr\_ok) is declared once such a flat tail is
established and spans at least a minimum length---15 responses
throughout this paper---the single structural constant of the
rank-stability test. This minimum plays the role of the earlier floor on
the number of observations required before evaluation and serves as a
monotone conservatism dial: a longer required tail fires later but is
more robust to a transient early plateau. The test introduces no slope
floor, no significance level, and no smoothing half-life; it is model
selection together with one length floor.

Concretely, the detector at response order \emph{n} proceeds in five
steps. (1) Compute the raw weighted Spearman correlation between the
cumulative share vectors at \emph{n} and \emph{n} $-$ \emph{w}, over the
union of domains observed in either window; domains missing from the
earlier window enter at share zero, ties receive average ranks, and
weights are the mean of the two share vectors, normalized to sum to one.
(2) For each candidate changepoint \emph{$\tau$} from 15 to \emph{n} $-$ 15,
fit the rising segment on the first $\tau$ observations by
pool-adjacent-violators and a constant to the remainder, and score the
model by BIC = \emph{{n}} $\cdot$ log(RSS/\emph{n}) + \emph{k} $\cdot$
log(\emph{n}), where \emph{k} counts the isotonic fit's distinct levels
plus two, one for the tail level and one for the changepoint location.
(3) Retain the best \emph{$\tau$} only if its BIC improves on a pure-rise
model scored the same way. (4) Certify the tail as flat only if a
constant model beats a linear model on the tail by BIC evaluated at the
autocorrelation-corrected effective sample size defined above. (5)
Declare rank stability at the first \emph{n} with a certified flat tail
of at least 15 responses; the verdict latches. Because both segments
must span at least 15 observations, the earliest possible firing is
response 30.

\begin{figure}[htbp]
\centering
\includegraphics[width=\textwidth,height=0.9\textheight,keepaspectratio]{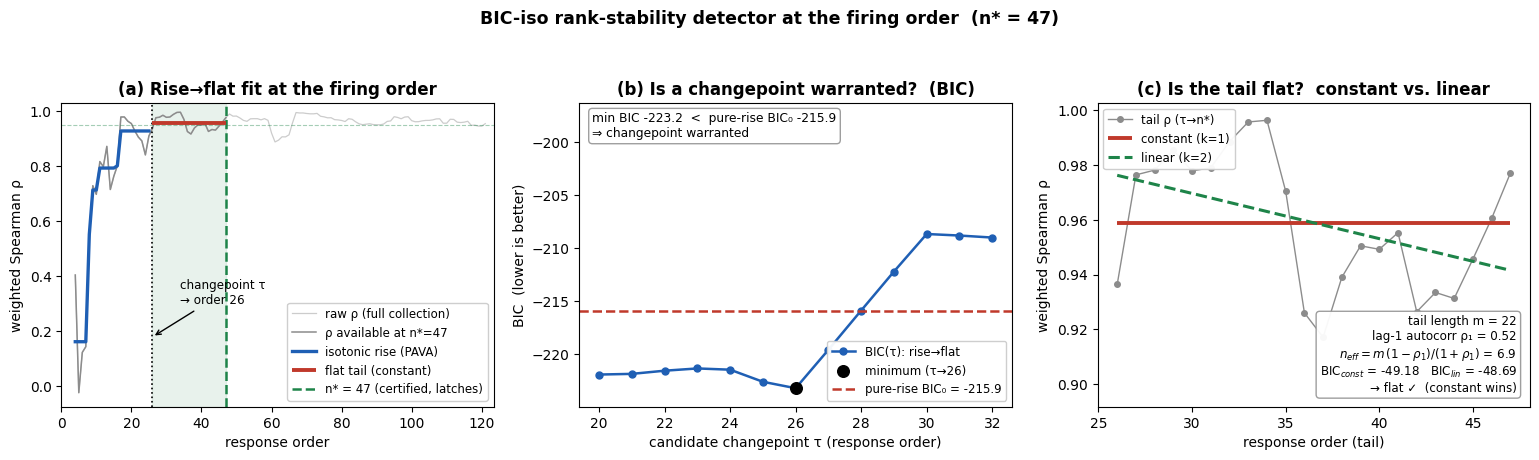}
\caption{BIC-iso rank stability detection for the ``coffee mugs'' topic
on SearchGPT. The series shows the weighted Spearman \emph{$\rho$} computed
at each response order between successive cumulative citation-share
windows. Isotonic regression (PAVA) is fit to the full \emph{$\rho$} history;
a two-segment model---a monotone-rising phase followed by a flat
tail---is preferred over a pure-rise model when it achieves a lower BIC.
$\tau$ denotes the BIC-selected plateau-onset index: the earliest response
order at which the correlation series transitions from its rising phase
to a sustained flat tail. Rank stability is declared at $\tau$ provided the
flat tail spans at least 15 consecutive responses and passes a flatness
test corrected for window autocorrelation. Once declared, the verdict
latches.}
\label{fig:3}
\end{figure}

Because the test locates a plateau rather than a level crossing, it
certifies stability at whatever correlation the platform's citation
density admits: a plateau at 0.88 is certified the same as one at 0.94.
This is the property a fixed threshold lacks. Once certified, the
verdict latches---rank stability is a terminal, one-way decision in the
sequential-analysis sense \citep{wald1947}, and the slow residual creep of
the cumulative correlation toward its ceiling is asymptotic approach,
not renewed re-ordering. Throughout this paper, we use a threshold-based
criterion (\emph{$\rho$} $\ge$ 0.95) as a reference benchmark; the contrast
between the two is itself an empirical demonstration of where fixed
thresholds fail. For period-over-period drift detection, a collection
whose rank trajectory has plateaued provides a principled baseline for
future checkpoint comparisons, regardless of the absolute correlation
level achieved.

\hypertarget{rank-uncertainty-quantification}{%
\subsection{Rank Uncertainty
Quantification}\label{rank-uncertainty-quantification}}

Even after rank order has stabilized, individual domain ranks carry
uncertainty that a point estimate does not capture. Rank confidence
intervals give the plausible range of rank positions for a domain across
repeated samples. Dense rank within platform is computed at each
bootstrap iteration over the domain set, and the resulting distribution
yields a CI on rank position. Because rank is discrete, these intervals
are interpreted as a range of positions rather than a continuous
estimate. Rank CIs inherit the reliability limitations described in
\S3.3: in the \emph{k} $\approx$ 5--20 citation danger zone, the underlying share
bootstrap distributions remain skewed and rank CIs derived from them may
misrepresent coverage.

\hypertarget{structural-signal-to-noise-ratio}{%
\subsection{Structural Signal-to-Noise
Ratio}\label{structural-signal-to-noise-ratio}}

Structural SNR operationalizes the question of whether the spread of
citation shares across the established set (domains whose share CI
excludes zero) is large relative to the precision of individual share
estimates.
\[\text{SNR}_{\text{est}} = \frac{\sigma(s_{d}:d \in D_{\text{est}})}{{\overline{h}}_{\text{CI}}}\]
where \emph{D}\textsubscript{est} is the established domain set,
\emph{s}\textsubscript{d} the citation share of established domain
\emph{d}, and \emph{U}\textsubscript{d} and \emph{L}\textsubscript{d}
the upper and lower bounds of its 95\% bootstrap confidence interval.
The per-domain half-width and its mean across the established set are
\[h_{\text{CI},d} = \frac{U_{d} - L_{d}}{2},\quad\quad{\overline{h}}_{\text{CI}} = \frac{1}{|D_{\text{est}}|}\sum_{d \in D_{\text{est}}}^{}h_{\text{CI},d}\]
The numerator is the standard deviation of established-domain citation
shares (the signal); the denominator is the mean half-width of their
citation-share confidence intervals (the noise).

When SNR \textless{} 1, the typical uncertainty around
established-domain shares is larger than the cross-domain spread, so the
observed ordering is dominated by estimation uncertainty. When SNR $\ge$ 1,
the share distribution carries enough structure, at the set level, to
support rank-based reporting at the top of the distribution, though it
does not imply that every adjacent rank is resolved. We treat structural
SNR as a practical inferential sufficiency heuristic rather than a
theoretically derived criterion, and the SNR = 1 threshold as an
operational default rather than a statistically canonical one. Its
motivation is geometric: the question is whether the spread of shares
across established domains is large relative to how precisely those
shares are measured. If the shares are tightly clustered, even narrow
CIs may not resolve the rank ordering; if the shares are spread out,
even moderate CI widths are sufficient. The criterion adapts to this
automatically because both of its terms are estimated from the same
accumulating sample: the spread of established shares and the per-domain
measurement uncertainty it is weighed against both come from the data in
hand, so the criterion sets its own bar rather than testing against a
target fixed in advance. The validation analysis in \S4.6 confirms
empirically that at this threshold the criterion accepts precisely the
orderings that agree with the full-window result and refuses precisely
those that agree least. In the sequential analysis framing, the
criterion belongs to the family of relative-precision stopping rules:
rather than requiring CI width to fall below a fixed target (as in
fixed-width procedures), it requires CI width to fall below a fraction
of the signal standard deviation, making the stopping target adaptive to
the observed share spread.

A compact scaling argument explains why structural SNR rises predictably
with collection and why it is self-calibrating. The established set is
defined by a citation-share confidence interval that excludes zero, and
that test requires the response-level bootstrap: the analytic Wilson
lower bound is strictly positive for any cited domain and so cannot
distinguish a share from zero, whereas the bias-corrected and
accelerated (BCa) bootstrap lower bound can reach zero. Membership
therefore rests on the bootstrap, and so does the CI half-width that
enters the SNR denominator.

The scaling itself does not depend on the interval's closed form. A
citation share is a ratio statistic over clustered citations; under
independent citations its variance would be
\emph{$\theta$}(1$-$\emph{$\theta$})/\emph{M}, where \emph{M} is the total citation
count; within-response citation clustering inflates this by the design
effect DEFF \citep{kish1965}, the ratio of the jackknife variance of the
share to its simple-random-sampling variance, so the effective citation
count is \emph{M}/DEFF and the standard error scales as
1/$\sqrt{\phantom{x}}$(\emph{M}/DEFF). Because \emph{M} grows in proportion to the
citation-bearing response count \emph{R}, the mean CI half-width across
the established set falls as 1/$\sqrt{\phantom{x}}$\emph{R}, while the signal \emph{$\sigma$}, the
standard deviation of established shares, is a property of the
distribution's shape and changes slowly once rank order has stabilized.
Structural SNR, the ratio of the two, therefore grows as $\sqrt{\phantom{x}}$\emph{R}, and
setting it to its target gives the minimum-response projection \emph{R}*
= \emph{R} $\cdot$ (SNR\_target / SNR)\textsuperscript{2}. A flatter distribution (smaller
\emph{$\sigma$}) requires more data than a steep one, and a platform that
accrues effective observations slowly, whether through sparse citation
or heavy within-response clustering, requires more responses to drive
the half-width down. Because the BCa interval has no closed-form
half-width, the honest object of the derivation is this rate, not a
plug-in sample-size equation.

That distinction governs how the projection can be used. \emph{R}*
resembles a classical fixed-width sample-size formula, but it cannot be
evaluated at the outset to set a collection budget. The signal \emph{$\sigma$}
is not stationary: as \emph{R} grows and intervals tighten, more domains
clear the zero boundary and enter the established set, reshaping
\emph{$\sigma$} and pulling the projected target downward, so \emph{R}* settles
only as \emph{$\sigma$} settles, near convergence rather than before it. DEFF
is itself estimated from the responses already collected, and the
projection is denominated in citation-bearing responses, which a
variable query-to-citation yield does not convert into a query budget in
advance. \emph{R}/SNR\textsuperscript{2} is therefore a confirmation signal that becomes
trustworthy only in the final stretch before convergence, not a
forecast: the criterion reports when enough data has accumulated, not
how much will ultimately be required. When a decision fixes an absolute
precision target (distinguishing a domain at 12\% share from one at
10\%, say, implies a target CI half-width near one percentage point),
the analyst can simply collect until the half-width on the domains of
interest falls below it; structural SNR is the self-calibrating default
for the common case in which no such target can be stated in advance.

The structural SNR, by contrast, is computed from a snapshot of the
established set at each response order: its numerator,
\emph{$\sigma$}(established shares), and its denominator, mean CI half-width,
both shift discretely whenever the established-set boundary moves or a
new response accrues. Without smoothing, a single-response fluctuation
could temporarily push SNR above the threshold and falsely declare
sufficiency, or a transient dip could cancel a declaration that should
persist. An exponentially weighted moving average (\emph{$\alpha$}, with a
half-life that adapts to collection size: max(3, 0.10 $\cdot$ n) responses,
equivalent to a smoothing constant near 0.2 early in collection and near
0.06 at n = 125) is therefore applied to the raw SNR series before the
sufficiency\_ok flag is evaluated. The smoothed series suppresses
threshold crossings caused by discrete jumps in the established set
while remaining responsive to genuine directional change. This smoothed
series is what governs the sufficiency\_ok flag and what is plotted in
the Results section.

The visual test is direct: compare the length of a typical error bar
(the mean CI half-width, drawn as the red arrow) to the half-height of
the blue band (the signal \emph{$\sigma$}). When SNR \textless{} 1 the arrow
extends beyond the half-band, when SNR = 1 the two match exactly, and
when SNR \textgreater{} 1 the band extends beyond the arrow.

\begin{figure}[htbp]
\centering
\includegraphics[width=\textwidth,height=0.9\textheight,keepaspectratio]{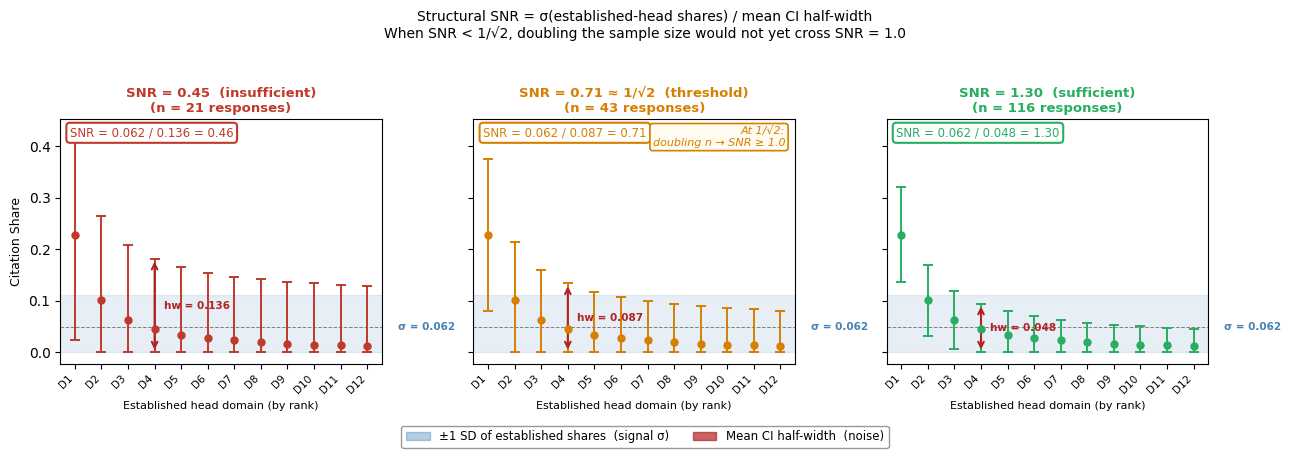}
\caption{Structural SNR = \emph{$\sigma$}(established shares) / mean CI
half-width, illustrated conceptually at three collection sizes using
synthetic domains (D1--D12). The blue band marks $\pm$1 SD of established
shares (signal \emph{$\sigma$} = 0.062 in all panels, held constant). The red
arrow marks the mean CI half-width (noise). Left panel (\emph{n} = 21,
SNR = 0.45): noise exceeds signal; the error bar extend beyond the blue
band. Center panel (\emph{n} = 43, SNR = 0.71): below the SNR = 1
threshold; signal does not yet match noise. Right panel (\emph{n} = 116,
SNR = 1.30): signal clearly exceeds noise. SNR grows slightly faster
than $\sqrt{n}$ at small \emph{n} because the interval's finite-sample
correction relaxes as \emph{n} grows; pure $1/\sqrt{n}$ scaling from SNR
= 0.45 at \emph{n} = 21 would give 1.06 at \emph{n} = 116, while the
exact value is 1.30.}
\label{fig:4}
\end{figure}

\begin{figure}[htbp]
\centering
\includegraphics[width=\textwidth,height=0.9\textheight,keepaspectratio]{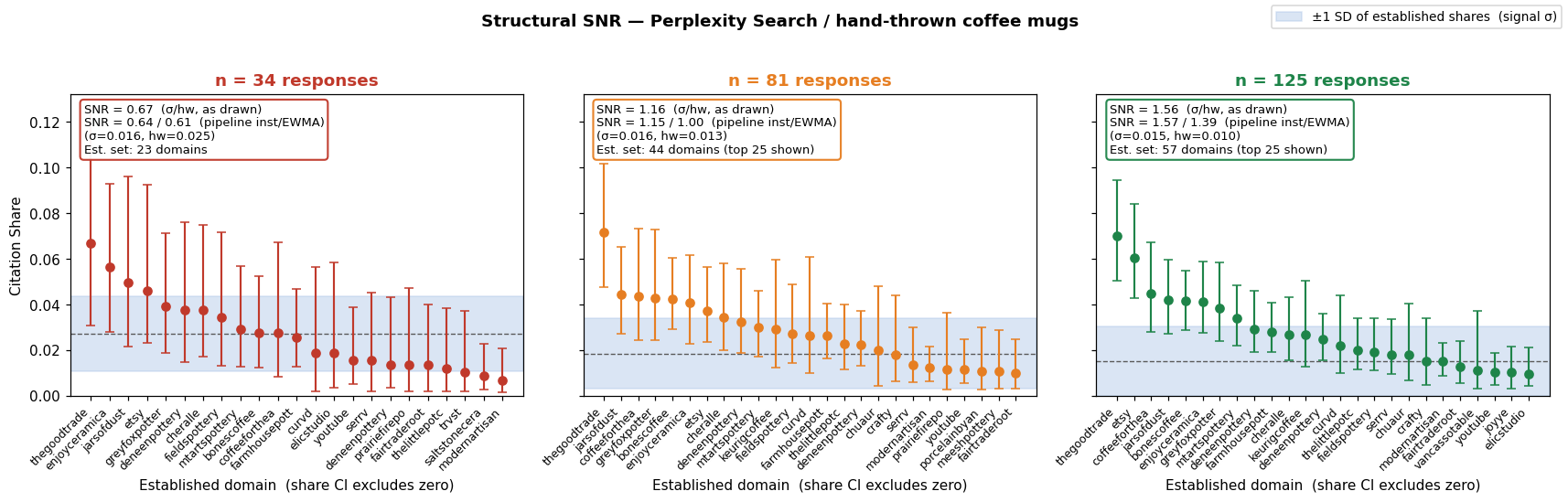}
\caption{Structural SNR using actual bootstrap CIs for the Perplexity /
``coffee mugs'' platform-topic combination at three collection sizes.
Each point is a domain citation share with 95\% bootstrap CI; the blue
band marks $\pm$1 SD of the established domain shares. SNR annotations show
the ratio and its components (\emph{$\sigma$}, hw).}
\label{fig:5}
\end{figure}

Figure~\ref{fig:5} illustrates the test with real data. At \emph{n} = 34 (SNR =
0.61, EWMA-smoothed, the series the sufficiency gate thresholds), when
rank\_corr\_ok is first determined, most CI half-widths extend well
beyond the $\pm$1 SD band. Very little statistical separation exists among
the established domains. At \emph{n} = 81 (SNR = 1.00), when
sufficiency\_ok is determined, the number of established domains has
increased to 44, and the CI bands of the top-ranked domains are
increasingly differentiated from the dense middle. At \emph{n} = 125
(SNR = 1.39), the end of the data gathering process, the CI bands are
perceptibly narrower. In the dense middle, the signal is often greater
than the noise, but adjacent domains still have overlapping CIs.
\emph{$\sigma$} barely changes across panels (0.016 $\rightarrow$ 0.016 $\rightarrow$ 0.015) because
the true share distribution is stable; SNR improves almost entirely
because hw falls as \emph{n} grows, which is precisely what the metric
is designed to capture. Between \emph{n} = 81 and \emph{n} = 125 the
mean half-width falls faster than the $1/\sqrt{n}$ per-domain rate alone
would predict, because newly established domains enter with narrow
absolute intervals and pull the set average down.

Both criteria share a common design principle. Rank stability requires
no target correlation level: the BIC-iso test detects a plateau
regardless of whether it occurs at \emph{$\rho$} = 0.88 or \emph{$\rho$} = 0.98.
Structural SNR requires no target CI width or precision level: the
threshold SNR $\ge$ 1 emerges from the relationship between the observed
signal and noise. Both stopping decisions are derived from the data's
own structure: the share spread of the established set, the empirical CI
widths, and the plateau geometry of the rank-correlation trajectory. The
framework specifies no target query count, correlation level, or CI
width. What remains externally set are two structural constants: the
15-response minimum plateau length, and the \emph{$\alpha$} smoothing applied
to the SNR signal, whose half-life scales with collection size, max(3,
0.10 $\cdot$ \emph{n}) responses. Neither encodes a target the data is asked
to reach; each governs how much confirmation the test demands before it
fires. The 15-response minimum also serves as a data-maturity floor for
the sufficiency test: structural SNR is not evaluated before 15
responses, because near-degenerate bootstrap intervals at very small
samples can spuriously inflate the ratio. Sufficiency firing orders of
15 in the Results are therefore censored at this floor.

\hypertarget{results}{%
\section{Results}\label{results}}

\hypertarget{dataset-and-distributional-structure}{%
\subsection{Dataset and Distributional
Structure}\label{dataset-and-distributional-structure}}

The analysis covers ten topics spanning diverse market structures and
information environments: ``marathon training shoes'', ``identity
protection and fraud prevention'', ``smoke detectors'', ``healthy meal
delivery for busy professionals'', ``collagen protein'', ``creamy milk
candy'', ``elevation-based corneal tomography'', ``hand-thrown coffee
mugs'', ``unusual travel destinations'', and ``ingredients for homemade
chili''. All analyses cover three platforms: Gemini, SearchGPT, and
Perplexity.

Table~\ref{tab:1} summarizes citation-bearing response counts, citation counts,
domain counts, and mean citations per citation-bearing response across
all platform-topic combinations. For Gemini and Perplexity, virtually
every submitted query yields a citation-bearing response; for SearchGPT,
a topic-dependent fraction of queries return no citations, so the
response count reflects only queries that produced at least one cited
source. Citation density differs substantially across platforms and, to
a lesser extent, across topics. Gemini produces between 19.9 and 50.1
citations per response depending on topic; Perplexity produces between
14.5 and 42.2; and SearchGPT produces between 4.2 and 6.3.
Within-response clustering compounds the density gap: the design effect
(DEFF, the ratio of jackknife to simple-random-sampling variance for a
domain's share), averaged across established domains at end of window,
runs 2.9--5.3 on Gemini (median 3.9), 3.7--7.8 on Perplexity (median
4.3), and 1.3--4.0 on SearchGPT (median 2.1). The effective response
count \emph{R}/DEFF, the number of independent-citation-equivalent
responses, therefore ends the window near a quarter of the nominal
response count on Gemini and near half on SearchGPT. These differences
drive every downstream result about rank stability timelines, rank CI
width, and the sensitivity of period-over-period tests.

{\footnotesize\setlength{\tabcolsep}{4pt}
\begin{longtable}[]{@{}lllllll@{}}
\caption{Dataset summary by topic and platform. Cit/Resp = mean citations per response.}\label{tab:1}\\
\toprule
Topic & Platform & Responses & Citations & Domains & Cit/Resp & Mean
DEFF\tabularnewline
\midrule
\endfirsthead
\toprule
Topic & Platform & Responses & Citations & Domains & Cit/Resp & Mean
DEFF\tabularnewline
\midrule
\endhead
Coffee mugs & Gemini & 125 & 3811 & 371 & 30.5 & 3.8\tabularnewline
& Perplexity & 125 & 2117 & 137 & 16.9 & 3.7\tabularnewline
& SearchGPT & 121 & 693 & 153 & 5.7 & 2.1\tabularnewline
Collagen protein & Gemini & 125 & 4985 & 385 & 39.9 & 4.0\tabularnewline
& Perplexity & 125 & 2627 & 213 & 21.0 & 4.0\tabularnewline
& SearchGPT & 121 & 725 & 170 & 6.0 & 2.1\tabularnewline
Corneal tomography & Gemini & 125 & 3785 & 254 & 30.3 &
3.6\tabularnewline
& Perplexity & 122 & 2224 & 91 & 18.2 & 3.8\tabularnewline
& SearchGPT & 117 & 494 & 50 & 4.2 & 1.7\tabularnewline
Creamy milk candy & Gemini & 125 & 2488 & 435 & 19.9 &
2.9\tabularnewline
& Perplexity & 125 & 1812 & 217 & 14.5 & 3.8\tabularnewline
& SearchGPT & 119 & 556 & 161 & 4.7 & 1.3\tabularnewline
Healthy meal delivery & Gemini & 124 & 4290 & 291 & 34.6 &
3.9\tabularnewline
& Perplexity & 125 & 2184 & 154 & 17.5 & 4.3\tabularnewline
& SearchGPT & 124 & 640 & 123 & 5.2 & 1.5\tabularnewline
Homemade chili & Gemini & 113 & 5656 & 356 & 50.1 & 4.1\tabularnewline
& Perplexity & 121 & 3956 & 180 & 32.7 & 4.8\tabularnewline
& SearchGPT & 104 & 594 & 156 & 5.7 & 2.1\tabularnewline
Identity protection & Gemini & 125 & 4516 & 544 & 36.1 &
5.3\tabularnewline
& Perplexity & 125 & 2820 & 320 & 22.6 & 7.8\tabularnewline
& SearchGPT & 104 & 660 & 230 & 6.3 & 2.9\tabularnewline
Marathon training shoes & Gemini & 125 & 3763 & 276 & 30.1 &
4.0\tabularnewline
& Perplexity & 124 & 2347 & 114 & 18.9 & 4.8\tabularnewline
& SearchGPT & 123 & 673 & 139 & 5.5 & 1.9\tabularnewline
Smoke detectors & Gemini & 122 & 4922 & 439 & 40.3 & 4.0\tabularnewline
& Perplexity & 125 & 5270 & 184 & 42.2 & 7.8\tabularnewline
& SearchGPT & 124 & 705 & 48 & 5.7 & 4.0\tabularnewline
Travel destinations & Gemini & 122 & 3597 & 496 & 29.5 &
3.5\tabularnewline
& Perplexity & 125 & 1907 & 190 & 15.3 & 4.4\tabularnewline
& SearchGPT & 113 & 704 & 190 & 6.2 & 2.7\tabularnewline
\bottomrule
\end{longtable}}

One structural feature of SearchGPT deserves mention. Unlike Gemini and
Perplexity, SearchGPT produces a non-negligible fraction of responses
containing no citations. In the ``identity protection'' topic, for
example, SearchGPT returned 104 citation-bearing responses from a query
batch in which Gemini and Perplexity each returned 125, a zero-citation
rate of approximately 17\%. The practical consequence is that the
response counts in Table~\ref{tab:1} for SearchGPT understate the collection
effort required to accumulate a given number of usable observations: a
collection budget of 125 queries may yield substantially fewer than 125
citation-bearing responses. Convergence response orders for SearchGPT
should therefore be read as sufficient citation-bearing responses, not
sufficient queries submitted. Practitioners can account for this by
inflating the query budget: if \emph{R} citation-bearing responses are
required and the platform's observed yield rate is \emph{Y} (the
fraction of submitted queries returning at least one citation), the
required query budget is \emph{N}\textsubscript{submit} = $\lceil$\emph{R} /
\emph{Y}$\rceil$. For a target of 100 citation-bearing responses at \emph{Y} $\approx$
0.83, this implies submitting approximately 121 queries rather than 100.

\begin{figure}[htbp]
\centering
\includegraphics[width=\textwidth,height=0.9\textheight,keepaspectratio]{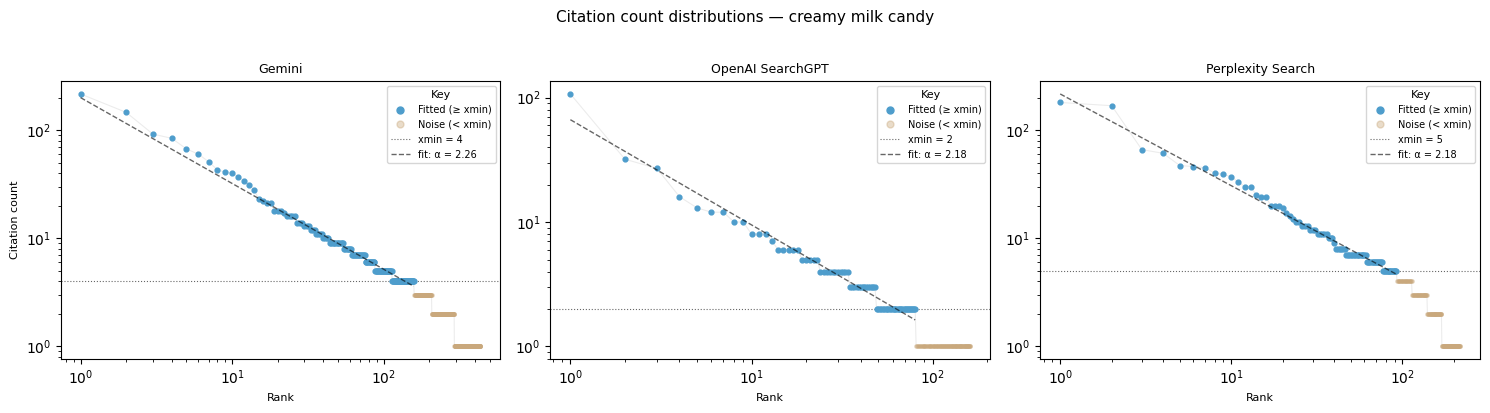}
\caption{Citation count distributions in log-log scale for the ``creamy
milk candy'' topic at full sample size, all three platforms. The dotted
horizontal line marks \emph{x}\textsubscript{min}; blue points are
domains that belong to the structured body of the citation distribution
(count $\ge$ \emph{x}\textsubscript{min}); tan points fall below that
boundary.}
\label{fig:6}
\end{figure}

Citation count distributions are broadly consistent with a power-law
model across the 29 platform-topic combinations for which fitting is
feasible. Figure~\ref{fig:6} shows one topic, ``creamy milk candy,'' on log-log
axes. Citation count declines approximately linearly with rank---the
characteristic signature of a power law---and the dotted line marks
\(x_{\min}\), the fitted boundary between the structured body of the
citation distribution and its sparse lower-ranked tail. Domains above
\(x_{\min}\) (blue) participate in the stable structure that defines the
ranking, whereas those below it (tan) occur too infrequently to follow
the same regular pattern.

Table~\ref{tab:2} reports the power-law fit parameters and the Heaps' Law exponent
for every platform-topic combination in the dataset. We fit each
distribution with the KS-minimized discrete maximum-likelihood procedure
of \citet{clauset2009}, which also supplies the semiparametric
bootstrap goodness-of-fit test reported there.

{\footnotesize\setlength{\tabcolsep}{4pt}
\begin{longtable}[]{@{}lllllll@{}}
\caption{Distributional structure by topic and platform. Domains = total distinct domains observed; \emph{x}\textsubscript{min} and \emph{$\alpha$} are the lower bound and exponent of the KS-minimized discrete power-law fit (Clauset et al., 2009); GOF \emph{p} is the semiparametric bootstrap goodness-of-fit \emph{p}-value, with values above 0.10 consistent with a power-law model; \emph{$\beta$} is the fitted Heaps' Law exponent for domain accumulation. Dashes (---) in the \emph{x}\textsubscript{min}, \emph{$\alpha$}, and GOF \emph{p} columns indicate fewer than 50 domains, below the floor required for a stable fit; \emph{$\beta$} does not require this floor.}\label{tab:2}\\
\toprule
\textbf{Topic} & \textbf{Platform} & \textbf{Domains} &
\textbf{\emph{x}\textsubscript{min}} & \emph{\textbf{$\alpha$}} & \textbf{GOF
\emph{p}} & \emph{\textbf{$\beta$}}\tabularnewline
\midrule
\endfirsthead
\toprule
\textbf{Topic} & \textbf{Platform} & \textbf{Domains} &
\textbf{\emph{x}\textsubscript{min}} & \emph{\textbf{$\alpha$}} & \textbf{GOF
\emph{p}} & \emph{\textbf{$\beta$}}\tabularnewline
\midrule
\endhead
Coffee mugs & Gemini & 371 & 4 & 1.95 & 0.51 & 0.61\tabularnewline
& Perplexity & 137 & 6 & 1.89 & 0.21 & 0.52\tabularnewline
& SearchGPT & 153 & 2 & 2.02 & 0.20 & 0.66\tabularnewline
Collagen protein & Gemini & 385 & 6 & 2.03 & 0.30 & 0.60\tabularnewline
& Perplexity & 213 & 5 & 2.03 & 0.41 & 0.60\tabularnewline
& SearchGPT & 170 & 1 & 1.84 & 0.089 & 0.69\tabularnewline
Corneal tomography & Gemini & 254 & 4 & 1.82 & 0.69 &
0.59\tabularnewline
& Perplexity & 91 & 2 & 1.56 & 0.26 & 0.49\tabularnewline
& SearchGPT & 50 & 1 & 1.84 & 0.021 & 0.54\tabularnewline
Creamy milk candy & Gemini & 435 & 4 & 2.26 & 0.99 & 0.69\tabularnewline
& Perplexity & 217 & 5 & 2.18 & 0.49 & 0.66\tabularnewline
& SearchGPT & 161 & 2 & 2.18 & 0.51 & 0.67\tabularnewline
Healthy meal delivery & Gemini & 291 & 4 & 1.82 & 0.044 &
0.57\tabularnewline
& Perplexity & 154 & 8 & 2.05 & 0.76 & 0.59\tabularnewline
& SearchGPT & 123 & 1 & 1.92 & 0.37 & 0.64\tabularnewline
Homemade chili & Gemini & 356 & 3 & 1.78 & 0.37 & 0.62\tabularnewline
& Perplexity & 180 & 8 & 1.90 & 0.74 & 0.64\tabularnewline
& SearchGPT & 156 & 1 & 1.85 & 0.25 & 0.73\tabularnewline
Identity protection & Gemini & 544 & 4 & 2.04 & 0.68 &
0.67\tabularnewline
& Perplexity & 320 & 4 & 2.08 & 0.12 & 0.68\tabularnewline
& SearchGPT & 230 & 3 & 2.32 & 0.29 & 0.75\tabularnewline
Marathon training shoes & Gemini & 276 & 3 & 1.81 & 0.10 &
0.59\tabularnewline
& Perplexity & 114 & 6 & 1.89 & 0.88 & 0.54\tabularnewline
& SearchGPT & 139 & 2 & 1.94 & 0.12 & 0.63\tabularnewline
Smoke detectors & Gemini & 439 & 6 & 2.06 & 0.019 & 0.62\tabularnewline
& Perplexity & 184 & 17 & 2.02 & 0.53 & 0.58\tabularnewline
& SearchGPT & 48 & --- & --- & --- & 0.57\tabularnewline
Travel destinations & Gemini & 496 & 4 & 2.08 & 0.42 &
0.68\tabularnewline
& Perplexity & 190 & 4 & 2.01 & 0.66 & 0.63\tabularnewline
& SearchGPT & 190 & 4 & 2.42 & 0.12 & 0.75\tabularnewline
\bottomrule
\end{longtable}}

Fitted \emph{$\alpha$} values range from 1.56 (Perplexity / ``corneal
tomography'') to 2.42 (SearchGPT / ``travel destinations''), with most
combinations between 1.78 and 2.26. The goodness-of-fit test of Clauset
et al. (2009) fails to reject the power-law model
(\emph{p}~\textgreater~0.10) for 25 of the 29 valid fits. The four
exceptions are Gemini / ``smoke detectors'' (\emph{p}~=~0.019),
SearchGPT / ``corneal tomography'' (\emph{p}~=~0.021), Gemini /
``healthy meal delivery'' (\emph{p}~=~0.044), and SearchGPT / ``collagen
protein'' (\emph{p}~=~0.089); these are treated as approximate rather
than ruled out as power-law distributed. One combination warrants
individual note: SearchGPT / ``corneal tomography'' has only 50 total
domains. With \emph{x}\textsubscript{min}~=~1, the fitted noise tier is
empty: all 50 domains fall within the structured body of the fit. The
established set is far smaller (12 domains at full sample, Table~\ref{tab:7}),
because establishment is defined by a share CI that excludes zero, not
by the fit. The power-law fit is technically valid but does not become
so until response 114, when the domain count finally reaches the
50-domain floor the KS fit requires. The established-set criterion does
not depend on that fit: structural sufficiency fires early (response 15,
the first order at which sufficiency is evaluated; the few established
shares being well separated), so rank stability is the binding condition
and the combination converges at \emph{n}*~=~61, gated by rank stability
rather than precision.

The Heaps' Law analysis confirms that new domains enter the observed
distribution throughout collection at a decelerating rate for every
platform-topic combination, with fitted exponents (\emph{$\beta$}, Table~\ref{tab:2})
consistent with the theoretical prediction from the corresponding Zipf
fits. This confirms that the continued appearance of new tail domains is
the structural consequence of the power-law distribution rather than
evidence of non-stationarity, and it explains the first-appearance
latency asymmetry: domains that ultimately rank highest appear earliest
in collection because they have the highest citation probabilities under
the Zipfian process, while tail domains continue appearing throughout
the 125-response window.

\hypertarget{rank-stability-during-collection-1}{%
\subsection{Rank Stability During
Collection}\label{rank-stability-during-collection-1}}

Figure~\ref{fig:7} shows weighted Spearman rank correlation trajectories for the
``travel destinations'' topic as a representative example, separately
for Gemini, Perplexity, and SearchGPT. Every platform-topic combination
begins with low correlation and rises as citations accumulate,
confirming that rank order is genuinely unstable during early collection
and only becomes interpretable as established domains separate from one
another. Table~\ref{tab:3} reports plateau-test results for all 30 combinations.

Table~\ref{tab:3} reports three stability benchmarks for each platform-topic
combination: the first response order at which raw \emph{$\rho$} $\ge$ 0.95 is
reached (\textbf{\emph{n}\textsubscript{r}}), the first response order
at which raw \emph{$\rho$} $\ge$ 0.95 was sustained for 15 consecutive responses
(\textbf{\emph{n}\textsubscript{s}}), and the response order at which
the BIC-iso plateau test first declares rank stability (\emph{n}*). The
BIC-iso test evaluates each possible changepoint as the start of a
candidate terminal window: the trajectory before the changepoint is fit
as a monotone rise, and the trajectory after the changepoint is tested
as a flat tail. BIC selects whether this two-part model improves on a
pure monotone-rise model, and stability is declared only when the
selected flat tail spans at least 15 responses and passes the flatness
check described in \S3.5.

In all 30 cases, the rank correlation reaches \emph{$\rho$} $\ge$ 0.95, often
early in the process (mean response order = 19.9). However, rank
correlation is jittery (as seen in Figure~\ref{fig:7}) and can fall below the
threshold within a few responses. To avoid spurious results, we can
require that the test's condition hold for 15 consecutive observations,
the same duration used in the BIC-iso plateau test. However, only 19
platform-topic combinations sustain \emph{$\rho$} $\ge$ 0.95 for 15 consecutive
observations, and they take much longer to do so (mean response order =
69.2).

In contrast, the BIC-iso test declares stability in all 30 combinations.
This contrast---19 combinations reaching sustained stability versus all
30 reaching the BIC-iso plateau---demonstrates that ceiling effects are
the primary limitation of any fixed-threshold approach: 11 combinations
never sustain 15 consecutive observations above \emph{$\rho$} = 0.95 not
because their ordering is still changing, but because their citation
density does not support reaching that level. When citation density is
low, each response cites only a handful of distinct domains, and the set
of cited domains changes substantially from query to query. Rank
correlation measures agreement in relative ordering across responses;
when few domains are shared between any two responses, the ordering is
inherently noisy, and rank correlation will not consistently reach 0.95
regardless of whether an underlying stable ordering exists. The plateau
test addresses this directly by detecting flatness rather than level.
Moreover, the BIC-iso test fires much earlier (mean response order =
40.0); in 18 of the 19 cases where the BIC-iso and sustained threshold
tests were both satisfied, BIC-iso fires at an earlier response order.
(The only exception was SearchGPT / ``corneal tomography''.)

{\footnotesize\setlength{\tabcolsep}{4pt}
\begin{longtable}[]{@{}lllllll@{}}
\caption{Response order at which raw \emph{$\rho$} $\ge$ 0.95 first reached (\textbf{\emph{{n}}\textsubscript{r}}), response order at which raw \emph{$\rho$} $\ge$ 0.95 was sustained for 15 periods (\textbf{\emph{{n}}\textsubscript{s}}), response order at which the rank stability criterion (rank\_corr\_ok) first fired (\emph{n}*), rank correlation at that point (\emph{$\rho$}*), and rank correlation at end of collection (\emph{$\rho$}\textsubscript{end}). Shaded cells identify \emph{$\rho$} $\ge$ 0.95.}\label{tab:3}\\
\toprule
\textbf{Topic} & \textbf{Platform} & \textbf{\emph{n}\textsubscript{r}}
& \textbf{\emph{n}\textsubscript{s}} & \textbf{\emph{n}*} &
\textbf{\emph{$\rho$}*} & \textbf{\emph{$\rho$}\textsubscript{end}}\tabularnewline
\midrule
\endfirsthead
\toprule
\textbf{Topic} & \textbf{Platform} & \textbf{\emph{n}\textsubscript{r}}
& \textbf{\emph{n}\textsubscript{s}} & \textbf{\emph{n}*} &
\textbf{\emph{$\rho$}*} & \textbf{\emph{$\rho$}\textsubscript{end}}\tabularnewline
\midrule
\endhead
Coffee mugs & Gemini & 13 & --- & 75 & 0.983 & 0.960\tabularnewline
& Perplexity & 20 & 51 & 34 & 0.937 & 0.979\tabularnewline
& SearchGPT & 17 & 80 & 47 & 0.977 & 0.955\tabularnewline
Collagen protein & Gemini & 21 & 60 & 33 & 0.954 & 0.972\tabularnewline
& Perplexity & 31 & 56 & 38 & 0.902 & 0.944\tabularnewline
& SearchGPT & 24 & --- & 37 & 0.921 & 0.963\tabularnewline
Corneal tomography & Gemini & 15 & 76 & 43 & 0.968 &
0.946\tabularnewline
& Perplexity & 10 & 50 & 34 & 0.928 & 0.972\tabularnewline
& SearchGPT & 8 & 43 & 61 & 0.996 & 0.981\tabularnewline
Creamy milk candy & Gemini & 20 & --- & 40 & 0.947 &
0.970\tabularnewline
& Perplexity & 29 & --- & 48 & 0.989 & 0.908\tabularnewline
& SearchGPT & 13 & 50 & 37 & 0.973 & 0.966\tabularnewline
Healthy meal delivery & Gemini & 19 & 72 & 33 & 0.976 &
0.942\tabularnewline
& Perplexity & 16 & 95 & 35 & 0.899 & 0.928\tabularnewline
& SearchGPT & 15 & 48 & 37 & 0.989 & 0.968\tabularnewline
Homemade chili & Gemini & 22 & 95 & 35 & 0.981 & 0.943\tabularnewline
& Perplexity & 17 & --- & 37 & 0.996 & 0.964\tabularnewline
& SearchGPT & 18 & --- & 33 & 0.991 & 0.954\tabularnewline
Identity protection & Gemini & 26 & --- & 34 & 0.876 &
0.976\tabularnewline
& Perplexity & 36 & --- & 42 & 0.861 & 0.920\tabularnewline
& SearchGPT & 32 & --- & 35 & 0.913 & 0.911\tabularnewline
Marathon training shoes & Gemini & 23 & 66 & 34 & 0.954 &
0.972\tabularnewline
& Perplexity & 27 & 76 & 43 & 0.948 & 0.982\tabularnewline
& SearchGPT & 19 & 50 & 39 & 0.980 & 0.934\tabularnewline
Smoke detectors & Gemini & 23 & 100 & 35 & 0.884 & 0.970\tabularnewline
& Perplexity & 15 & 109 & 33 & 0.963 & 0.928\tabularnewline
& SearchGPT & 4 & 40 & 33 & 0.997 & 0.969\tabularnewline
Travel destinations & Gemini & 29 & 98 & 41 & 0.928 &
0.952\tabularnewline
& Perplexity & 20 & --- & 49 & 0.970 & 0.928\tabularnewline
& SearchGPT & 26 & --- & 46 & 0.906 & 0.956\tabularnewline
\bottomrule
\end{longtable}}

The plateau test declares rank stability for all 30 platform-topic
combinations within the 125-response collection window. Joint rank
stability (all three platforms simultaneously stable) is achieved for
all ten topics, from ``smoke detectors'' at response 35 to ``coffee
mugs'' at response 75. Stability firing orders fall within a moderate
band: every \emph{n}* falls between 33 and 75 responses. As \S4.5 shows,
the wide variation in total data requirements comes from the sufficiency
condition, not from stability.

Two patterns stand out. First, the platform ordering of convergence
speed varies across topics. No platform is consistently fastest or
slowest: each of the three platforms has at least one topic on which it
plateaus earliest, with \emph{n}* = 33, and at least one on which it
plateaus latest, sometimes by a substantial margin, as in Gemini /
``coffee mugs'' (\emph{n}* = 75) and SearchGPT / ``corneal tomography''
(\emph{n}* = 61). The same platforms that converge earliest for some
topics are latest for others. This cross-topic variation indicates that
convergence speed is jointly determined by topic-level market structure
and platform citation behavior, rather than by platform identity alone.
Second, the stability orders occupy a narrower band (33 to 75) than the
conjunctive orders in \S4.5, which range from 33 to 94 and include three
combinations that never converge. Although stability fires last in most
converged combinations (16 of 27, \S4.5), it never fires later than
response 75; the long delays and all three convergence failures come
from sufficiency.

These results establish the practical implication directly: no fixed
query count can be justified across all topics and platforms. The
BIC-iso plateau test resolves this by detecting a plateau rather than a
level---all 30 combinations reach stability between \emph{n}* = 33 and
\emph{n}* = 75, while the sustained \emph{$\rho$} $\ge$ 0.95 benchmark leaves 11
of the 30 unsatisfied within the collection window. Stability, however,
is only half the criterion. \S4.5 shows that structural sufficiency is
the binding constraint in 11 of the 27 converging combinations and is
never achieved in three.

\hypertarget{rank-uncertainty-at-full-sample-size}{%
\subsection{Rank Uncertainty at Full Sample
Size}\label{rank-uncertainty-at-full-sample-size}}

Even where the ordering is stable, the precision of individual ranks
varies enormously across the distribution, and this pattern holds across
all 30 platform-topic combinations. Table~\ref{tab:4} reports median CI width
across the full domain vocabulary for each combination. At the low end,
SearchGPT / ``smoke detectors'' (48 domains) has a median CI of 25.5
rank positions; at the high end, Gemini / ``identity protection'' (544
domains) reaches 329.9 positions. Expressed as a fraction of total
vocabulary, median CI widths span 38--66\% with a median of 58\%. Table
5 pools the same widths by reporting scope: among each combination's ten
most-cited domains the median CI spans 5.0 rank positions (90th
percentile 14.1, with 18.7\% exceeding ten positions); across the
established tier it spans 69.1; across the full vocabulary, 131.0. Rank
uncertainty persists well inside the reporting-relevant portion of the
distribution. These large medians reflect the many tail domains whose
rank assignments are essentially arbitrary, but even within the
established set, CI widths widen substantially toward the noise
boundary. A rank ordering can be fully stable and still be too imprecise
at the lower portion of the established set to support meaningful
competitive analysis. This is the gap the structural SNR criterion in
\S3.7 closes: it identifies whether the top of the established set is
measured precisely enough to trust the ordering.

{\small
\begin{longtable}[]{@{}lllll@{}}
\caption{Rank CI width statistics at full sample size \emph{(n} = 125 responses, 95\% CI). Median CI width and percentage of domains with CI width exceeding 10 rank positions are shown by topic and platform.}\label{tab:4}\\
\toprule
Topic & Platform & Domains & Median rank CI width & \% with width
\textgreater{} 10\tabularnewline
\midrule
\endfirsthead
\toprule
Topic & Platform & Domains & Median rank CI width & \% with width
\textgreater{} 10\tabularnewline
\midrule
\endhead
Coffee mugs & Gemini & 371 & 217.0 & 96\tabularnewline
& Perplexity & 137 & 65.0 & 92\tabularnewline
& SearchGPT & 153 & 95.0 & 92\tabularnewline
Collagen protein & Gemini & 385 & 199.0 & 95\tabularnewline
& Perplexity & 213 & 106.1 & 91\tabularnewline
& SearchGPT & 170 & 102.6 & 94\tabularnewline
Corneal tomography & Gemini & 254 & 134.6 & 93\tabularnewline
& Perplexity & 91 & 35.0 & 84\tabularnewline
& SearchGPT & 50 & 27.5 & 74\tabularnewline
Creamy milk candy & Gemini & 435 & 262.1 & 96\tabularnewline
& Perplexity & 217 & 132.0 & 93\tabularnewline
& SearchGPT & 161 & 104.1 & 93\tabularnewline
Healthy meal delivery & Gemini & 291 & 156.7 & 94\tabularnewline
& Perplexity & 154 & 80.0 & 92\tabularnewline
& SearchGPT & 123 & 76.0 & 89\tabularnewline
Homemade chili & Gemini & 356 & 213.0 & 94\tabularnewline
& Perplexity & 180 & 102.0 & 89\tabularnewline
& SearchGPT & 156 & 98.5 & 93\tabularnewline
Identity protection & Gemini & 544 & 329.9 & 97\tabularnewline
& Perplexity & 320 & 186.2 & 95\tabularnewline
& SearchGPT & 230 & 151.0 & 96\tabularnewline
Marathon training shoes & Gemini & 276 & 153.0 & 92\tabularnewline
& Perplexity & 114 & 54.3 & 86\tabularnewline
& SearchGPT & 139 & 82.0 & 94\tabularnewline
Smoke detectors & Gemini & 439 & 227.0 & 94\tabularnewline
& Perplexity & 184 & 98.4 & 87\tabularnewline
& SearchGPT & 48 & 25.5 & 71\tabularnewline
Travel destinations & Gemini & 496 & 303.6 & 96\tabularnewline
& Perplexity & 190 & 110.1 & 91\tabularnewline
& SearchGPT & 190 & 116.5 & 96\tabularnewline
\bottomrule
\end{longtable}}

\begin{figure}[htbp]
\centering
\includegraphics[width=\textwidth,height=0.9\textheight,keepaspectratio]{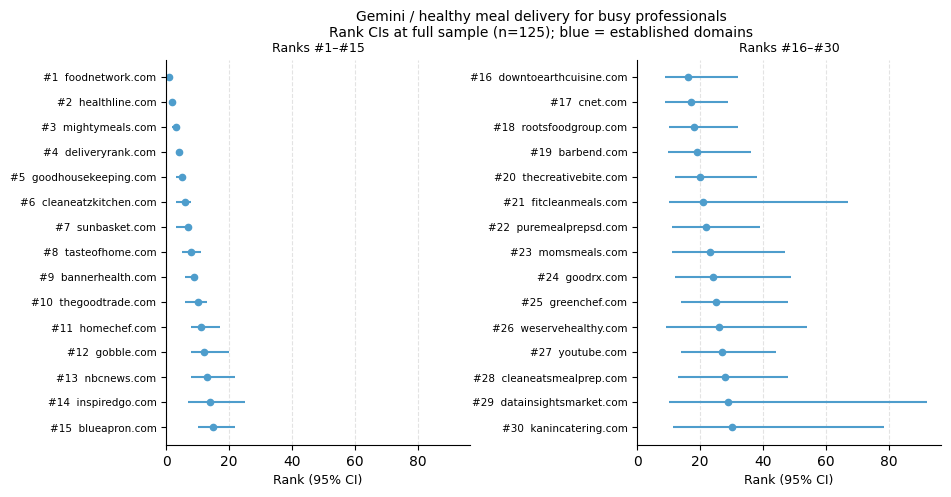}
\caption{Rank CI forest plot for Gemini / ``healthy meal delivery'' at
full sample size \emph{(n} = 125). Each horizontal bar is the 95\%
bootstrap rank CI for one domain. The top-ranked domains have narrow CIs
and are distinguishable, but adjacent domains in the middle of the
distribution have overlapping intervals spanning 30--80+ rank positions.}
\label{fig:8}
\end{figure}

Figure~\ref{fig:8} illustrates this gradient directly. At the top of the
established set, CIs are narrow: the first eight or nine domains are
clearly distinguishable from one another, and rank-based reporting is
reliable. Deeper in the distribution, where established domains share
similar citation frequencies, CIs widen substantially, and adjacent
ranks become statistically indistinguishable. By ranks \#26--\#30,
several CI widths span as many as 80 rank positions, making the apparent
ordering unreliable even though all these domains are established. Rank
matters; the question is whether you have collected enough data to know
which ranks can be trusted.

{\small
\begin{longtable}[]{@{}lllll@{}}
\caption{Rank CI width by reporting scope, pooled across all 30 platform-topic combinations at full sample size (95\% CIs). Top 10 = the ten most-cited domains per combination; established = domains whose citation-share CI excludes zero; widths in rank positions.}\label{tab:5}\\
\toprule
Scope & Domains & Median rank CI width & 90th percentile width & \% with
width \textgreater{} 10\tabularnewline
\midrule
\endfirsthead
\toprule
Scope & Domains & Median rank CI width & 90th percentile width & \% with
width \textgreater{} 10\tabularnewline
\midrule
\endhead
Top 10 domains & 300 & 5.0 & 14.1 & 18.7\tabularnewline
Established domains & 1,693 & 69.1 & 217.2 & 84.2\tabularnewline
Full vocabulary & 7,067 & 131.0 & 327.0 & 93.3\tabularnewline
\bottomrule
\end{longtable}}

\hypertarget{structural-sufficiency}{%
\subsection{Structural Sufficiency}\label{structural-sufficiency}}

Rank stability certifies that the ordering of citation shares has
stopped changing; it does not certify that the shares are measured
precisely enough to be informative. Structural sufficiency addresses
that second condition. It asks whether the structure that remains, the
spread of citation shares across the established set, is large enough
relative to the uncertainty in those shares to support inference. When
structural SNR $\ge$ 1, the spread of citation shares across the established
set is large relative to a typical confidence interval, meaning the
ordering reflects a real signal rather than estimation noise. Where
structural SNR never reaches 1, additional collection would be
required---or, in some cases, the platform-topic combination may not
support reliable share estimation within any feasible collection window.

\begin{figure}[htbp]
\centering
\includegraphics[width=\textwidth,height=0.9\textheight,keepaspectratio]{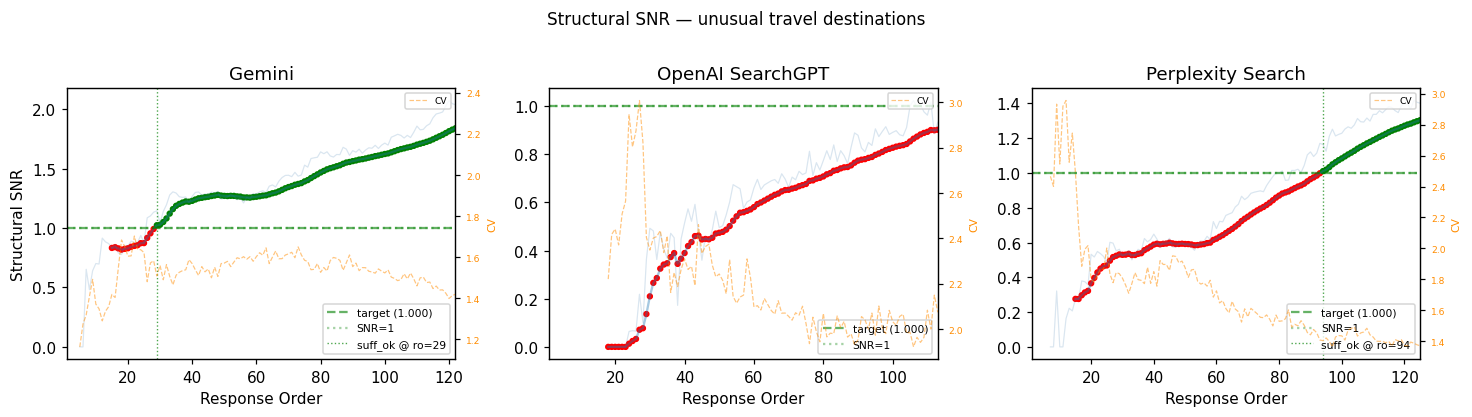}
\caption{Structural SNR trajectories for the ``travel destinations''
topic, shown separately for each platform. The Structural SNR increases
slowly as more data are gathered and the mean CI of established domains
decreases (yellow dotted line). The horizontal dashed line marks the
SNR\,=\,1 sufficiency threshold, and the dotted vertical line indicates
the response order at which the EWMA-smoothed SNR crosses the threshold.
The points are colored red prior to sufficiency\_ok and green
thereafter. Table~\ref{tab:6} reports the structural SNR for all 30 platform-topic
combinations.}
\label{fig:9}
\end{figure}

Figure~\ref{fig:9} shows the structural SNR trajectories for the ``travel
destinations'' topic across all three platforms. The trajectories
illustrate the range of outcomes. Gemini crosses the SNR $\ge$ 1 threshold
at \emph{n} = 29 responses, before its rank stability fires at \emph{n}*
= 41; sufficiency is not the binding constraint here. Perplexity crosses
at \emph{n} = 94, after rank stability at \emph{n}* = 49, so sufficiency
is the binding constraint in this case. SearchGPT achieves rank
stability at \emph{n}* = 46, but the structural SNR never crosses the
threshold within the collection window, and the combination does not
converge.

Across all 30 platform-topic combinations, structural sufficiency is the
binding constraint in 11 of the 27 that converge: rank stability fires
first in these combinations, and the conjunction must wait for
structural SNR~$\ge$~1. Three combinations do not converge within the
125-response collection window, all on SearchGPT. For ``identity
protection'' and ``travel destinations'', the end-of-window structural
SNR is 0.74 and 0.90 respectively. For ``homemade chili'', structural
SNR ends the window at 0.89. In each signal-limited case the established
citation shares are distributed too narrowly---relative to the width of
the confidence intervals---for the criterion to fire within the
125-response collection window.

\hypertarget{conjunctive-convergence-results}{%
\subsection{Conjunctive Convergence
Results}\label{conjunctive-convergence-results}}

The preceding sections treat rank stability (\S4.2), rank uncertainty
(\S4.3), and structural sufficiency (\S4.4) as separate analyses. Their
conjunction defines the paper's central convergence criterion: a
platform-topic combination is declared sufficient when both
rank\_corr\_ok (the BIC-iso plateau test on the rank correlation
trajectory) and sufficiency\_ok (structural SNR $\ge$ 1) hold
simultaneously. Table~\ref{tab:6} reports, for each combination, the first
response order at which each condition fired and the first response
order at which the conjunction held; the conjunctive order is in every
case the later of the two component orders. The criterion converges for
27 of 30 combinations within the 125-response collection window. Median
conjunctive convergence orders are 42.0 for Gemini, 37.0 for SearchGPT
(across its seven converging combinations), and 51.0 for Perplexity. The
two conditions share the binding role: structural SNR fires last in 11
of the 27 converged combinations, and rank stability fires last in the
other 16. Sufficiency is not a formality. Where sufficiency binds, it
delays the stop by anywhere from a few responses to 47 responses (e.g.,
in Perplexity / ``coffee mugs'', rank stability fires at response 34 but
structural sufficiency does not fire until response 81). In three
combinations, all on SearchGPT, structural sufficiency never fires
within the window (\S4.4).

{\footnotesize\setlength{\tabcolsep}{4pt}
\begin{longtable}[]{@{}llllll@{}}
\caption{Conjunctive convergence by platform-topic combination: first firing of rank stability (\emph{n}*), first firing of structural sufficiency (\emph{n}\textsubscript{SNR}), first firing of the conjunction (\emph{n}\textsubscript{conj}), and the EWMA-smoothed structural SNR at end of window (SNR\textsubscript{end}). The conjunctive order equals the later of \emph{n}* and \emph{n}\textsubscript{SNR} in every converged case. Dashes (---) indicate that the test criterion was not met within the 125-response collection window. Values of \emph{n}\textsubscript{SNR} = 15 coincide with the first response order at which sufficiency is evaluated (\S3.7) and are censored at that floor.}\label{tab:6}\\
\toprule
Topic & Platform & \emph{n}* & \emph{n}\textsubscript{SNR} &
\emph{n}\textsubscript{conj} & SNR\textsubscript{end}\tabularnewline
\midrule
\endfirsthead
\toprule
Topic & Platform & \emph{n}* & \emph{n}\textsubscript{SNR} &
\emph{n}\textsubscript{conj} & SNR\textsubscript{end}\tabularnewline
\midrule
\endhead
Coffee mugs & Gemini & 75 & 60 & 75 & 1.59\tabularnewline
& Perplexity & 34 & 81 & 81 & 1.39\tabularnewline
& SearchGPT & 47 & 34 & 47 & 1.64\tabularnewline
Collagen protein & Gemini & 33 & 32 & 33 & 2.12\tabularnewline
& Perplexity & 38 & 58 & 58 & 1.61\tabularnewline
& SearchGPT & 37 & 24 & 37 & 2.22\tabularnewline
Corneal tomography & Gemini & 43 & 15 & 43 & 4.71\tabularnewline
& Perplexity & 34 & 15 & 34 & 3.62\tabularnewline
& SearchGPT & 61 & 15 & 61 & 5.77\tabularnewline
Creamy milk candy & Gemini & 40 & 27 & 40 & 2.09\tabularnewline
& Perplexity & 48 & 55 & 55 & 1.69\tabularnewline
& SearchGPT & 37 & 21 & 37 & 2.57\tabularnewline
Healthy meal delivery & Gemini & 33 & 15 & 33 & 3.06\tabularnewline
& Perplexity & 35 & 22 & 35 & 2.07\tabularnewline
& SearchGPT & 37 & 20 & 37 & 3.53\tabularnewline
Homemade chili & Gemini & 35 & 33 & 35 & 2.50\tabularnewline
& Perplexity & 37 & 30 & 37 & 2.35\tabularnewline
& SearchGPT & 33 & --- & --- & 0.89\tabularnewline
Identity protection & Gemini & 34 & 63 & 63 & 1.39\tabularnewline
& Perplexity & 42 & 74 & 74 & 1.32\tabularnewline
& SearchGPT & 35 & --- & --- & 0.74\tabularnewline
Marathon training shoes & Gemini & 34 & 45 & 45 & 2.40\tabularnewline
& Perplexity & 43 & 47 & 47 & 2.32\tabularnewline
& SearchGPT & 39 & 45 & 45 & 1.74\tabularnewline
Smoke detectors & Gemini & 35 & 55 & 55 & 1.66\tabularnewline
& Perplexity & 33 & 43 & 43 & 1.94\tabularnewline
& SearchGPT & 33 & 22 & 33 & 2.31\tabularnewline
Travel destinations & Gemini & 41 & 29 & 41 & 1.84\tabularnewline
& Perplexity & 49 & 94 & 94 & 1.30\tabularnewline
& SearchGPT & 46 & --- & --- & 0.90\tabularnewline
\bottomrule
\end{longtable}}

Shaded rows: not converged.

\begin{figure}[htbp]
\centering
\includegraphics[width=\textwidth,height=0.9\textheight,keepaspectratio]{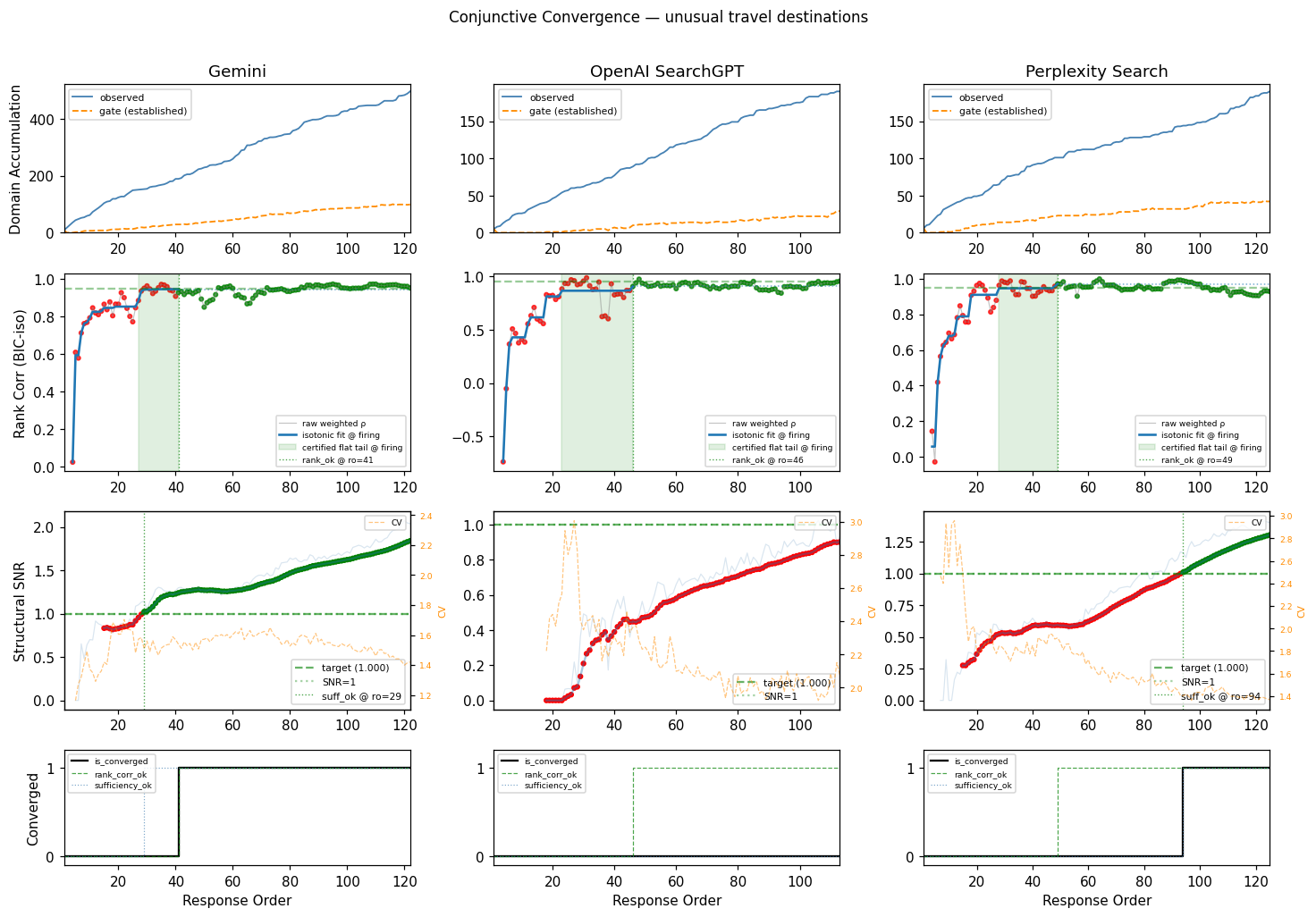}
\caption{Conjunctive convergence for the ``travel destinations'' topic
across Gemini, Perplexity, and SearchGPT. Top row: unique domain
accumulation. Second row: weighted Spearman rank correlation with
BIC-iso plateau fit. Third row: structural SNR (EWMA-smoothed) with the
sufficiency threshold at SNR\,=\,1. Bottom row: conjunctive convergence
flag, True when both rank\_corr\_ok and sufficiency\_ok hold
simultaneously. Conjunctive convergence orders: Gemini
\emph{n}\textsubscript{conj}= 41, Perplexity
\emph{n}\textsubscript{conj}=\,94. SearchGPT does not converge within
the 125-response collection window.}
\label{fig:10}
\end{figure}

Figure~\ref{fig:10} illustrates the conjunctive convergence process for the
``travel destinations'' topic. The four-row panel shows domain
accumulation, rank correlation, structural SNR, and the conjunctive
convergence flag together across all three platforms in a single view.
Gemini achieves conjunction at \emph{n} = 41: structural SNR clears 1 at
response 29, well ahead of rank stability at \emph{n}* = 41, so
stability is the binding constraint. Perplexity achieves rank stability
at \emph{n}* = 49, but structural SNR does not clear the threshold until
response 94, making sufficiency the binding constraint and setting the
conjunctive order to \emph{n} = 94. SearchGPT achieves rank stability at
\emph{n}* = 46 but does not cross the SNR threshold within the
collection window, and the combination does not converge.

The mechanism is the same in all three non-converging combinations: at
roughly six citations per response, SearchGPT's confidence intervals
cannot narrow enough within 125 responses to resolve a signal \emph{$\sigma$}
that is already small. More collection would eventually push structural
SNR past 1; the criterion's verdict is that 125 responses is not enough
here, which is precisely the platform-topic dependence that makes a
fixed universal budget untenable.

The zero-citation response rate noted in \S4.1 affects the interpretation
but not the validity of the convergence orders. The criterion asks
whether the observed citation data (however many citation-bearing
responses have been accumulated) is sufficient to support rank-based
reporting. That question is correctly answered in terms of
citation-bearing responses. The operational implication for collection
planning is separate: because SearchGPT returns no citations in a
topic-dependent fraction of queries, achieving a target number of
citation-bearing responses requires submitting more total queries than
the convergence order implies. The convergence criterion is not
misspecified; the query budget must account for platform yield.

\hypertarget{validation-do-stopped-orderings-persist}{%
\subsection{Validation: Do Stopped Orderings
Persist?}\label{validation-do-stopped-orderings-persist}}

The validation supports the criterion on both sides of its decision, and
its value is concentrated at the tail. The orderings it refuses include
the weakest agreements in the dataset, and the orderings it certifies
avoid the worst failures that stopping at rank stability alone would
have admitted. The convergence criterion certifies that an ordering has
stabilized and is measured precisely enough to act on. The central
question is whether stopping at the firing order actually yields the
ordering one would have obtained by collecting the full budget. The
check is direct. For each combination, we compare the cumulative
established-set ordering at the stopping point against the ordering at
the end of the 125-response window, using the same weighted Spearman
correlation as the stability criterion (weights proportional to average
shares) over the established domains at full sample (those whose
citation-share CI excludes zero), along with the overlap between the
top-10 lists. Because the two orderings share the data collected up to
the stopping point, this is an internal consistency check against the
best available estimate, not a comparison with external ground truth.
Fixing the established set at its full-sample membership likewise uses
information unavailable at the stopping point, so the agreement figures
are retrospective audits of the certified orderings, not quantities an
analyst could have computed at the stop. Stronger validation would
compare stopped orderings against independent repeated collections or
held-out responses; we leave that design to future work.

Three results stand out. First, stopping at rank stability alone is not
safe: agreement at \emph{n}* ranges from 0.364 to 0.986 (median 0.823)
across all 30 combinations, and the NC refusal with the weakest
ordering---SearchGPT / ``identity protection'' (0.364, the single
worst)---demonstrates exactly why the sufficiency condition is needed.
Perplexity / ``identity protection'' (0.713) and Gemini / ``identity
protection'' (0.722) also show weak agreement at \emph{n}*. This is
exactly why sufficiency is required as a second condition: the
criterion's refusals are vindicated by the data. Second, where
structural sufficiency delays the stop it usually helps and rarely
hurts, though the average gain is modest: agreement improved in all 11
combinations where structural SNR delayed the stop beyond \emph{n}*,
with most of the benefit concentrated in a few cases (Perplexity /
``identity protection'' from 0.713 to 0.866; Perplexity / ``travel
destinations'' from 0.842 to 0.978) and small elsewhere. Third, the
certified stops are sound rather than flawless: across the 27 converged
combinations, agreement at \emph{n}\textsubscript{conj} spans 0.717 to
0.986 with a median of 0.844, and the median top-10 overlap is 80\%, so
on average two of the top ten domains still differ from the full-window
ranking. The criterion's contribution shows up at the floor: it lifts
the worst certified agreement to 0.717, well above the 0.364 that rank
stability alone would have admitted.

{\footnotesize\setlength{\tabcolsep}{4pt}
\begin{longtable}[]{@{}lllllll@{}}
\caption{Agreement between stopped and end-of-window established-set orderings: weighted Spearman correlation evaluated at the rank-stability firing order (\emph{n}*) and at the conjunctive firing order (\emph{n}\textsubscript{conj}). A dash (---) indicates that the test criterion was not met within the 125-response collection window. The established set is fixed at the full-sample established tier (domains whose citation-share CI excludes zero). \emph{d}\textsubscript{est} = number of domains with citation-share CI excluding zero at full sample.}\label{tab:7}\\
\toprule
\textbf{Topic} & \textbf{Platform} &
\textbf{\emph{d}\textsubscript{est}} & \emph{\textbf{n*}} &
\textbf{\emph{$\rho$} at \emph{n*}} & \textbf{\emph{n}\textsubscript{conj}} &
\textbf{\emph{$\rho$} at \emph{n}\textsubscript{conj}}\tabularnewline
\midrule
\endfirsthead
\toprule
\textbf{Topic} & \textbf{Platform} &
\textbf{\emph{d}\textsubscript{est}} & \emph{\textbf{n*}} &
\textbf{\emph{$\rho$} at \emph{n*}} & \textbf{\emph{n}\textsubscript{conj}} &
\textbf{\emph{$\rho$} at \emph{n}\textsubscript{conj}}\tabularnewline
\midrule
\endhead
Coffee mugs & Gemini & 88 & 75 & 0.947 & 75 & 0.947\tabularnewline
& Perplexity & 57 & 34 & 0.861 & 81 & 0.954\tabularnewline
& SearchGPT & 37 & 47 & 0.781 & 47 & 0.781\tabularnewline
Collagen protein & Gemini & 109 & 33 & 0.746 & 33 & 0.746\tabularnewline
& Perplexity & 66 & 38 & 0.791 & 58 & 0.821\tabularnewline
& SearchGPT & 45 & 37 & 0.837 & 37 & 0.837\tabularnewline
Corneal tomography & Gemini & 57 & 43 & 0.915 & 43 &
0.915\tabularnewline
& Perplexity & 31 & 34 & 0.909 & 34 & 0.909\tabularnewline
& SearchGPT & 12 & 61 & 0.986 & 61 & 0.986\tabularnewline
Creamy milk candy & Gemini & 81 & 40 & 0.802 & 40 & 0.802\tabularnewline
& Perplexity & 41 & 48 & 0.876 & 55 & 0.877\tabularnewline
& SearchGPT & 30 & 37 & 0.890 & 37 & 0.890\tabularnewline
Healthy meal delivery & Gemini & 98 & 33 & 0.833 & 33 &
0.833\tabularnewline
& Perplexity & 47 & 35 & 0.825 & 35 & 0.825\tabularnewline
& SearchGPT & 23 & 37 & 0.885 & 37 & 0.885\tabularnewline
Homemade chili & Gemini & 80 & 35 & 0.877 & 35 & 0.877\tabularnewline
& Perplexity & 34 & 37 & 0.885 & 37 & 0.885\tabularnewline
& SearchGPT & 25 & 33 & 0.818 & --- & ---\tabularnewline
Identity protection & Gemini & 122 & 34 & 0.722 & 63 &
0.838\tabularnewline
& Perplexity & 64 & 42 & 0.713 & 74 & 0.866\tabularnewline
& SearchGPT & 30 & 35 & 0.364 & --- & ---\tabularnewline
Marathon training shoes & Gemini & 71 & 34 & 0.822 & 45 &
0.839\tabularnewline
& Perplexity & 36 & 43 & 0.841 & 47 & 0.862\tabularnewline
& SearchGPT & 37 & 39 & 0.703 & 45 & 0.717\tabularnewline
Smoke detectors & Gemini & 142 & 35 & 0.722 & 55 & 0.817\tabularnewline
& Perplexity & 49 & 33 & 0.768 & 43 & 0.844\tabularnewline
& SearchGPT & 12 & 33 & 0.799 & 33 & 0.799\tabularnewline
Travel destinations & Gemini & 97 & 41 & 0.758 & 41 &
0.758\tabularnewline
& Perplexity & 42 & 49 & 0.842 & 94 & 0.978\tabularnewline
& SearchGPT & 30 & 46 & 0.716 & --- & ---\tabularnewline
\bottomrule
\end{longtable}}

Shaded rows: not converged.

This analysis also answers the post-firing decline objection raised by
Table~\ref{tab:3}, where several combinations fire with \emph{$\rho$}* above 0.95 and
end the window below it (Gemini / ``homemade chili'' declines from 0.981
to 0.943, Perplexity / ``smoke detectors'' from 0.963 to 0.928). The
declining quantity is the window-to-window correlation, which oscillates
as marginal citations shift positions near the threshold. The ordering
itself does not degrade: the certified orderings for those two
combinations agree with the end-of-window orderings at 0.88 and 0.84 by
the validation measure, with top-10 overlaps of 90\% and 80\%
respectively. Threshold jitter in \emph{$\rho$} is a property of the
monitoring statistic, not of the ranking being certified.

\hypertarget{sensitivity-to-the-sufficiency-threshold}{%
\subsection{Sensitivity to the Sufficiency
Threshold}\label{sensitivity-to-the-sufficiency-threshold}}

\begin{figure}[htbp]
\centering
\includegraphics[width=\textwidth,height=0.9\textheight,keepaspectratio]{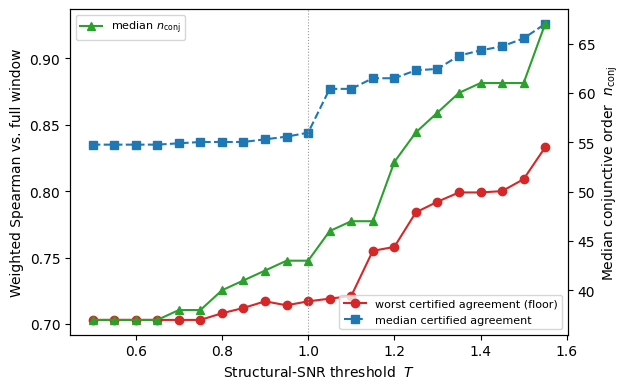}
\caption{Sensitivity of the conjunctive criterion to the
structural-SNR threshold \emph{T}, with the BIC-iso rank-stability test
held fixed. Left axis: weighted Spearman agreement between the certified
ordering and the full-window ordering, shown as the median across
certified combinations (squares) and the worst case, or floor (circles).
Right axis: median conjunctive convergence order (triangles). The dotted
line marks the default \emph{T} = 1.0. Agreement and collection cost
both rise smoothly with the threshold; below \emph{T} $\approx$ 0.8 agreement is
flat, so lower thresholds add no fidelity while admitting the weakest
orderings.}
\label{fig:11}
\end{figure}

The structural-SNR threshold of 1.0 is an operational default, so we
test how the criterion behaves as it moves. Holding the rank-stability
test fixed, we sweep the threshold \emph{T} from 0.5 to 1.5 and
recompute, at each value, how many of the 30 combinations converge
within the window, how long they take, and how closely the certified
ordering agrees with the full-window result (Figure~\ref{fig:11}). The result is a
smooth cost-fidelity tradeoff, not a fragile operating point. Below
\emph{T} $\approx$ 0.75 the certified orderings do not improve: every
combination converges, but the worst certified agreement stays pinned
near 0.70 and the median near 0.84, so a lower threshold only stops
sooner and admits the weakest orderings the data produce. From there the
floor rises steadily rather than plateauing (0.71 at 0.85, 0.72 at the
default, 0.76 by 1.2, and 0.80 by 1.4), as the weakest combinations
either accumulate enough data to sharpen or fall below the rising bar,
while the median agreement climbs (0.84 at \emph{T} = 1.0 to 0.89 by
1.2) and the median conjunctive order rises from 43 to 53 responses.
Beyond \emph{T} $\approx$ 1.3 combinations begin falling outside the
125-response budget, with 27 converging at 1.3, 25 at 1.35, and 23 by
1.4. \emph{T} = 1.0 certifies 27 of the 30 combinations at a median
agreement of 0.84 and a worst case of 0.717, and it carries the
geometric meaning that the spread of established shares equals their
measurement noise. The criterion is not fragile at T = 1.0: agreement
and collection cost vary smoothly as the threshold changes, so the
threshold can be treated as an operating choice rather than a brittle
constant. A practitioner who can afford more collection, or who needs a
tighter floor, can read a stricter operating point directly off the
curve.

\hypertarget{discussion}{%
\section{Discussion}\label{discussion}}

The conjunctive convergence criterion identifies a richer condition than
either rank stability or structural sufficiency alone, and the results
show both conditions doing real work. Rank stability can hold while
confidence intervals remain wide enough to make adjacent domains
statistically indistinguishable: structural SNR fired after stability in
11 of the 27 converged combinations and never fired in three more.
Structural SNR can clear its threshold while the rank ordering is still
settling: SNR fired before stability in the other 16, by margins as
large as 46 responses (SearchGPT / ``corneal tomography'': SNR at
response 15, stability at 61). The conjunction protects against both
failure modes: stability ensures the ordering is not changing, and
sufficiency ensures the ordering is precise enough to be meaningful.

The finding that citation density and within-query clustering---not
platform identity---govern convergence speed has a direct practical
implication. When it converges, SearchGPT does so at response counts
comparable to Gemini and Perplexity. The key is effective sample size,
not total citation volume: SearchGPT emits roughly one-sixth as many
citations per response, but its low within-query citation clustering
(median end-of-window DEFF 2.1) means each response contributes roughly
half an effective response, while Gemini's high clustering (median DEFF
3.9) means each response contributes roughly one-quarter (the effective
response count \emph{R}/DEFF of \S4.1). Equal-\emph{n} collection budgets
are therefore not equally informative across platforms, and
practitioners comparing platforms on the same nominal query count may be
comparing unequal amounts of effective data.

The ceiling-effect finding has a direct implication for collection
protocol design. Any platform with low citation density will appear
unstable under a fixed correlation threshold regardless of whether its
ordering has settled, because the mechanics of sparse citation make the
threshold unreachable rather than unmet. Practitioners relying on fixed
thresholds for platforms like SearchGPT---or any future platform with
similarly sparse citation behavior---will systematically over-collect.
The plateau detector resolves this by asking whether correlation has
stopped changing rather than whether it has reached a target level. The
practical upshot is that collection protocols should not be calibrated
to a fixed \emph{$\rho$} target; they should be calibrated to whether the
trajectory has flattened.

The SNR sufficiency criterion belongs to the family of
relative-precision stopping rules described by \citet{chow1965}.
Unlike fixed-width procedures that require an absolute precision target,
the SNR criterion sets the stopping target relative to the observed
signal spread \emph{$\sigma$}, making it self-calibrating to the citation
distribution at hand. When the practitioner has a specific precision
requirement---for instance, a target CI half-width of one percentage
point for a domain with a ten percent share---that target can be stated
directly. The SNR threshold is the appropriate default when no such
requirement can be specified in advance. A threshold-sensitivity
analysis (\S4.7) confirms the choice is not fragile: agreement and
collection cost vary smoothly across a wide range of the threshold, with
no value at which the criterion's behavior changes abruptly.

The validation analysis of \S4.6 provides an internal check on the
framework's central claim. A stopping rule is only as good as the
orderings it certifies, and the certified orderings agree reasonably
with the end-of-window orderings (median 0.844); more to the point, the
refusals include the weakest orderings in the dataset, which is where
the sufficiency condition earns its keep. The same analysis quantifies
the cost of dropping the sufficiency condition: rank stability alone
would have certified an ordering agreeing only 0.364 with the final
ordering on SearchGPT / ``identity protection''. Where structural
sufficiency delayed the stop, the certified ordering improved in all 11
cases.

The convergence criterion can serve as active stopping logic: once both
conditions hold simultaneously, the collection can be concluded and
reporting begun. In practice, the criterion is most naturally reviewed
at intervals consistent with balanced query design---for instance, after
each complete set of query types has been run---to preserve equal
representation across prompt types, user profiles, and consumer intent
categories. When the criterion is not satisfied at end of collection,
the appropriate response is to continue gathering data, guided by the
constraints of a balanced dataset.

\hypertarget{limitations}{%
\subsection{Limitations}\label{limitations}}

The dataset covers ten topics across three platforms during a single
collection period. Generalizability of the specific convergence
thresholds to other platforms, other topic categories, or different time
periods is not established empirically here. The qualitative
findings---that citation density and clustering govern convergence
speed, and that no fixed query count is universally sufficient---are
structural consequences of the citation-generation process and expected
to generalize.

Convergence is defined relative to the query population that is sampled,
not to a query-independent notion of visibility. The citation
distribution a platform produces depends on which queries are posed, so
a different query design (real user queries, a different generating
model, or a different mix of query types) would sample a different
distribution and could converge at a different point. The criterion
certifies that a given query design has been sampled sufficiently; it
does not certify that the design is itself representative of the
questions a brand cares about. That representativeness is an input the
analyst must supply.

The framework assumes the citation distribution is stationary over the
collection window. It asks whether enough data has been drawn from a
fixed distribution, not whether that distribution has itself changed.
Generative platforms are updated over time (models, indexes, and ranking
logic all change), and a collection that spans such a change would see
genuine movement in the citation distribution. The convergence criterion
cannot separate that movement from ordinary sampling variability; within
a single collection it would read real drift as continued instability or
as a failure to converge. Detecting change between collection periods is
a distinct problem, one this paper does not explore.

The analysis treats all queries within a platform-topic combination as a
single set. In practice, the queries used for measurement may contain
multiple facets, query types with options such as `expert
recommendations' or `product comparisons,' various geographies, and
more. An analyst focusing attention on a single facet option would
necessarily have a proportionally smaller effective response count
\emph{C} to work with, which widens bootstrap confidence intervals and
raises the CI half-width entering the SNR denominator. Rank stability
may likewise require more responses to certify on the reduced series.
The sufficiency criterion captures this shrinkage automatically: a lower
SNR on the filtered subset signals that the narrower dataset cannot yet
resolve citation-share differences at the required precision, and
additional queries of the target type are needed before conclusions can
be drawn. Analysts performing query-type-stratified comparisons should
treat convergence as a per-stratum property and report the sufficiency
ratio separately for each stratum.

Similarly, the analysis treats each platform-topic combination on its
own. Combining several platforms into a single, all-up measure
introduces a further complication, because platforms differ in citation
density by more than sixfold: Gemini and Perplexity emit tens of
citations per response while SearchGPT emits four to six. Pooling raw
citation counts across platforms weights each platform by its citation
volume, so a dense platform overshadows sparse ones and the aggregate
comes to reflect mostly the densest. Share-based metrics avoid this by
normalizing within each platform before the platforms are combined, so
that each contributes equally regardless of volume. The same discipline
must extend to the uncertainty estimates: confidence intervals, and
therefore structural SNR and rank CIs, should be computed per platform
and then averaged across platforms, not pooled from raw counts, since a
pooled bootstrap would once again let the high-volume platform dominate
the variance. Convergence, likewise, is best treated as a per-platform
property; an all-up result is ready only once every constituent platform
has individually converged.

Bootstrap BCa confidence intervals for rank CIs and rank displacement
CIs are computationally intensive at scale. For studies where bootstrap
cost is prohibitive, a Wilson score interval \citep{wilson1927} with
per-domain design effect adjustment (Wilson + DEFF) provides a fast
analytical alternative for the convergence criterion. However, an
important subtlety governs its use: a Wilson interval's lower bound is
strictly positive for any domain with at least one citation, so it
cannot reproduce the boundary between observed and established domains
used throughout this paper, which is defined by a confidence interval
that excludes zero. A Wilson-based pipeline must therefore identify the
established set against a positive reference, such as the noise floor
implied by the power-law fit, rather than against zero, and the
resulting set will differ slightly from the bootstrap's. All results
reported here use the bootstrap, whose lower bound can reach zero and so
identifies the established set directly.

\hypertarget{conclusion}{%
\section{Conclusion}\label{conclusion}}

We introduced a two-stage convergence framework for citation visibility
measurement in generative search engines. The first stage, rank
stability, determines when the rank ordering of observed domains,
dominated by high-share domains through weighting, has stopped changing.
The second stage, structural sufficiency, determines when citation share
estimates are precise enough relative to the distributional spread to
support meaningful rank-based inference. Both conditions are required;
neither alone is sufficient.

Applied to 30 platform-topic combinations across three platforms, the
conjunctive criterion converges for 27 within a 125-response collection
window. The three non-converging combinations are all on SearchGPT and
all are signal-limited, with citation shares too tightly packed relative
to measurement precision for structural SNR to clear its threshold
within 125 responses. The validation analysis supports the criterion on
both sides of the decision: the certified stops agree reasonably with
the end-of-window orderings, and, more tellingly, the three
signal-limited refusals include the worst-agreeing ordering in the
dataset. The framework is fully data-adaptive: it conditions the
stopping criterion on the observed citation density, within-query
clustering, and share spread at each evaluation point, rather than on
assumed parameter values. The stopping rule sets no target query count,
correlation level, or CI width, and the structural constants it does
retain govern how much confirmation it requires, not where it must end
up; the framework tells you when you have arrived, not how far away you
are. No fixed query count can be justified a priori across contexts; in
this dataset no budget below 94 responses would have certified every
converging combination, and three combinations were not certifiable
within 125. A collection that passes both conditions supports rank-based
reporting at the top of the distribution, share comparisons among
established domains, and a principled baseline for period-over-period
change detection. What the criterion certifies is that the ordering has
stabilized and that the established set as a whole has separated from
measurement noise, not that every adjacent pair of ranks is resolved.

\nocite{*}
\bibliographystyle{plainnat}
\bibliography{references}

@book{barlow1972,
  author    = {Barlow, R. E. and Bartholomew, D. J. and Bremner, J. M. and Brunk, H. D.},
  title     = {Statistical Inference Under Order Restrictions},
  publisher = {Wiley},
  year      = {1972}
}

@article{bartlett1946,
  author  = {Bartlett, M. S.},
  title   = {On the theoretical specification and sampling properties of autocorrelated time-series},
  journal = {Supplement to the Journal of the Royal Statistical Society},
  volume  = {8},
  number  = {1},
  pages   = {27--41},
  year    = {1946}
}

@article{chow1965,
  author  = {Chow, Y. S. and Robbins, Herbert},
  title   = {On the asymptotic theory of fixed-width sequential confidence intervals for the mean},
  journal = {Annals of Mathematical Statistics},
  volume  = {36},
  number  = {2},
  pages   = {457--462},
  year    = {1965}
}

@article{clauset2009,
  author  = {Clauset, Aaron and Shalizi, Cosma Rohilla and Newman, M. E. J.},
  title   = {Power-law distributions in empirical data},
  journal = {SIAM Review},
  volume  = {51},
  number  = {4},
  pages   = {661--703},
  year    = {2009}
}

@article{efron1987,
  author  = {Efron, Bradley},
  title   = {Better bootstrap confidence intervals},
  journal = {Journal of the American Statistical Association},
  volume  = {82},
  number  = {397},
  pages   = {171--185},
  year    = {1987}
}

@book{heaps1978,
  author    = {Heaps, H. S.},
  title     = {Information Retrieval: Computational and Theoretical Aspects},
  publisher = {Academic Press},
  year      = {1978}
}

@book{kish1965,
  author    = {Kish, Leslie},
  title     = {Survey Sampling},
  publisher = {Wiley},
  year      = {1965}
}

@article{sielinski2026,
  author  = {Sielinski, Ronald},
  title   = {Quantifying uncertainty in {AI} visibility: A statistical framework for generative search measurement},
  journal = {arXiv preprint},
  year    = {2026},
  note    = {arXiv:2603.08924}
}

@book{wald1947,
  author    = {Wald, Abraham},
  title     = {Sequential Analysis},
  publisher = {Wiley},
  year      = {1947}
}

@article{wilson1927,
  author  = {Wilson, Edwin B.},
  title   = {Probable inference, the law of succession, and statistical inference},
  journal = {Journal of the American Statistical Association},
  volume  = {22},
  number  = {158},
  pages   = {209--212},
  year    = {1927}
}

@book{zipf1949,
  author    = {Zipf, George Kingsley},
  title     = {Human Behavior and the Principle of Least Effort},
  publisher = {Addison-Wesley},
  year      = {1949}
}

@inproceedings{gao2023,
  author    = {Gao, Tianyu and Yen, Howard and Yu, Jiatong and Chen, Danqi},
  title     = {Enabling large language models to generate text with citations},
  booktitle = {Findings of the Association for Computational Linguistics: EMNLP 2023},
  year      = {2023}
}

@article{jarvelin2002,
  author  = {J{\"a}rvelin, Kalervo and Kek{\"a}l{\"a}inen, Jaana},
  title   = {Cumulated gain-based evaluation of {IR} techniques},
  journal = {ACM Transactions on Information Systems},
  volume  = {20},
  number  = {4},
  pages   = {422--446},
  year    = {2002}
}

@article{li2024,
  author  = {Li, Aleksandra and Sinnamon, Luanne},
  title   = {Generative {AI} search engines as arbiters of public knowledge: An audit of bias and authority},
  journal = {Proceedings of the Association for Information Science and Technology},
  volume  = {61},
  number  = {1},
  year    = {2024}
}

@article{moffat2008,
  author  = {Moffat, Alistair and Zobel, Justin},
  title   = {Rank-biased precision for measurement of retrieval effectiveness},
  journal = {ACM Transactions on Information Systems},
  volume  = {27},
  number  = {1},
  year    = {2008}
}

@article{mokander2023,
  author  = {M{\"o}kander, Jakob and Schuett, Jonas and Kirk, Hannah R. and Floridi, Luciano},
  title   = {Auditing large language models: A three-layered approach},
  journal = {AI \& Ethics},
  volume  = {4},
  pages   = {1017--1035},
  year    = {2023}
}

@inproceedings{vigna2015,
  author    = {Vigna, Sebastiano},
  title     = {A weighted correlation index for rankings with ties},
  booktitle = {Proceedings of the 24th International Conference on World Wide Web (WWW 2015)},
  year      = {2015}
}

@article{webber2010,
  author  = {Webber, William and Moffat, Alistair and Zobel, Justin},
  title   = {A similarity measure for indefinite rankings},
  journal = {ACM Transactions on Information Systems},
  volume  = {28},
  number  = {4},
  year    = {2010}
}

@article{zhang2025,
  author  = {Zhang, Peng and Ye, Qiang and Tang, Kun and Zhao, Wayne Xin and Wen, Ji-Rong},
  title   = {Source coverage and citation bias in {LLM}-based vs.\ traditional search engines},
  journal = {arXiv preprint},
  year    = {2025},
  note    = {arXiv:2512.09483}
}

\end{document}